\documentclass[review]{elsarticle}
\usepackage{hyperref}
\usepackage[T1]{fontenc}
\usepackage{amsmath}
\usepackage[latin1]{inputenc}
\usepackage{color}
\usepackage{amsfonts}

\usepackage{multirow}
\usepackage{longtable}

\journal{Astroparticle Physics}

\begin{document}

\def\diffnu{\left.\frac{dN_{\nu}}{dE_{\nu}}\right|_{\rm diffuse}}
\def\diffnures{\left.\frac{dN_{\nu}}{dE_{\nu}}\right|_{\rm diffuse}^{\rm res}}
\def\diffnuunres{\left.\frac{dN_{\nu}}{dE_{\nu}}\right|_{\rm diffuse}^{\rm unres}}

\begin{frontmatter}

\title{Gamma-ray emitting supernova remnants as the origin of Galactic cosmic rays?}

\author[bo]{Julia Becker Tjus\corref{mycorrespondingauthor}}
\ead{julia.tjus@rub.de}

\author[bo]{Bj\"orn Eichmann}

\author[bo]{Mike Kroll}

\author[bo]{Nils Nierstenh\"ofer}

\cortext[mycorrespondingauthor]{Corresponding author}

\address[bo]{Theoretische Physik IV: Plasma-Astroteilchenphysik\\Fakult\"at f\"ur Physik und Astronomie\\Ruhr-Universit\"at Bochum\\44780 Bochum\\Germany}

\begin{abstract}
The origin of cosmic rays is one of the long-standing mysteries in physics and astrophysics. Simple arguments suggest that a scenario of supernova remnants (SNRs) in the Milky Way as the dominant sources for the cosmic ray population below the knee could work: in a generic calculation, it can be shown that these objects can provide the energy budget necessary to explain the observed flux of cosmic rays. However, this argument is based on the assumption that all sources behave in the same way, i.e.\ they all have the same energy budget, spectral behavior and maximum energy. In this paper, we investigate if a realistic population of SNRs is capable of producing the cosmic ray flux as it is observed below the knee. We use 21 SNRs that are well-studied from radio wavelengths up to gamma-ray energies. It could be shown previously (Mandelartz \& Becker Tjus 2015) that the high-energy bump in the energy spectrum of these 21 sources can be dominated by hadronic emission. Here, gamma-rays are produced via $\pi^{0}-$decays from cosmic ray interactions in molecular clouds near the supernova remnant, which serves as the cosmic ray accelerator. The cosmic ray spectra show a large variety in their energy budget, spectral behavior and maximum energy. These sources are assumed to be representative for the total class of SNRs, where we assume that about 100 - 200 cosmic ray emitting SNRs should be present today. Finally, we use these source spectra to simulate the cosmic ray transport from individual SNRs in the Galaxy with the GALPROP code for cosmic ray propagation. We find that the cosmic ray budget can  be matched well for a diffusion coefficient that is close to $D\propto E^{0.3}$. A stronger dependence on the energy, e.g.\ $E^{0.5}$, would lead to a spectrum at Earth that is too steep when compared to what is detected and the energy budget cannot be matched, in particular toward high energies. 
We conclude that gamma-ray emitting SNRs can be a representative sample of cosmic ray emitting sources. In the future, experiments like CTA and HAWC will help to distinguish hadronic from leptonic sources and to further constrain the maximum energy of the sources. With better data, the uncertainties in this calculation will be reduced in order to develop even stronger conclusions on the question if SNRs can  be the sources of Galactic cosmic rays.
\end{abstract}

\begin{keyword}
Cosmic ray spectrum \sep Supernova Remnants \sep Cosmic ray propagation
\end{keyword}

\end{frontmatter}


\section{Introduction\label{intro:sec}}
The question of the origin of cosmic rays is one of the most crucial ones in physics and astrophysics. Since their first detection in 1912 \cite{hess1912}, the energy spectrum of cosmic rays has been studied in great detail, revealing a general power-law structure of the differential energy spectrum, $dN/dE\propto E^{\gamma}$ and distinct features as the two prominent breaks in the power-law at $10^{15}$~eV (the so-called cosmic ray knee) and $10^{18.5}$~eV (the so-called cosmic ray ankle), see e.g.\ reviews given in \cite{gaisser1991,stanev2003}. Due to the deflection of these charged particles in cosmic magnetic fields, the observed flux arrives at a very high level of isotropy and sources are difficult to identify. During the past years, great progress has been made to start tying the observed cosmic ray flux to specific source classes. In particular, gamma-rays and neutrinos, arising from cosmic ray interactions in the vicinity of their production region, could be used to start to identify the cosmic ray origin. These neutral particles can arise in photohadronic interactions or in the interaction of cosmic rays with matter,
\begin{eqnarray}
p\,p&\rightarrow& \pi^{\pm,0}\\
p\,\gamma &\rightarrow& \Delta^{+}\rightarrow \left\{\begin{array}{l}p\,\pi^{0}\\n\,\pi^{+}\end{array}\right.\\
\pi^{+}/\pi^{-}&\rightarrow& (e^{+}+\nu_{e}+\overline{\nu}_{\mu})+\nu_{\mu}/(e^{-}+\overline{\nu}_{e}+\nu_{\mu})+\overline{\nu}_{\mu}\\
\pi^{0}&\rightarrow& \gamma\gamma\,.
\end{eqnarray}
Here, it is assumed that all pions and muons decay before further interaction/acceleration in the source.

The detection of astrophysical high-energy gamma-rays, neutrinos and charged particles has provided us with important pieces of information on Galactic cosmic ray sources in the past few years:
\begin{enumerate}
\item  {\bf Gamma-rays} are in general considered an ambiguous signature, as both bremsstrahlung and Inverse Compton scattering can contribute to a potential signal from $\pi^{0}-$decays, see e.g.\ \cite{blumenthal_gould1970,schlickeiser2002}. There are, however, two quite unambiguous signatures which make it possible to identify the $\pi^{0}-$decay energy spectrum: (1) The high-energy cutoff at above $100$~TeV photon energy corresponds to acceleration up to $1$~PeV primary energy, assuming that about $10\%$ of the primary's energy is transferred into photons. With leptonic processes, it is extremely difficult to reach such high energies. Future observations with HAWC and CTA, improving the sensitivity of gamma-ray detection above $100$~TeV will help finding those sources accelerating up to PeV-energies. (2) The low-energy cutoff in the $\pi^{0}-$decay spectrum is basically independent of the power-law index of the primary spectrum and is located at around $200$~MeV photon energy. Recently, it was possible to identify such a pronounced low-energy cutoff at $100$~MeV  for the two SNRs W44 \cite{abdo_w44} and IC443 \cite{adbo_ic443}, using the Fermi satellite. While these detections provide first proof that SNRs actually do accelerate cosmic rays, these two sources have steep energy spectra, or even an energy cutoff in the TeV-range. The unambiguous identification of PeVatrons in the Galaxy must be done in the future at the highest energies. Today, about 30 gamma-ray spectra from SNRs have been detected with around 20 shell-type SNR detections with imaging air Cherenkov Telescopes at TeV energies \cite{tevcat}. Although the cutoff at the highest energies could in many cases not be identified yet, the spectral behavior up to 10 TeV gamma-ray energy is known. This corresponds to cosmic ray proton energies of around 100 TeV, reaching to a factor of 10 below the knee. Thus, current spectra can now be used to try to estimate a possible contribution of SNRs to the total Galactic cosmic ray spectrum and energy budget.
\item Astrophysical  {\bf neutrinos} provide unambiguous proof of hadronic interactions, but their detection is challenging \cite{halzen_klein2010}. The first experimental proof of the existance of astrophysical high-energy neutrinos succeeded with the 1~km$^{3}$-sized IceCube detector \cite{icecube2013,icecube2014}. The detection in the neutrino energy range corresponds to cosmic ray energies between approximately $600$~TeV up to $>40$~PeV, assuming that about $1/20$th of the primary's energy is going into the neutrino. In a dedicated analysis of three years of data, 37 neutrinos were detected at a background of $\sim 4-6$~atmospheric neutrinos per year. The approximate number of $6-10$ astrophysical neutrinos per year does not point towards one or a few sources, but represents a diffuse flux from a larger number of sources. From the spatial distribution of events, it is clear that a larger fraction of the flux must come from sources off the Galactic plane. Different estimates show that the diffuse neutrino flux from Galactic sources can only contribute with $2-4$ events or less \cite{kistler_beacom2006,ahlers_murase2014,neronov2014,winter_galactic2014,kachelriess2014,mandelartz2015}. Future measurements with IceCube and IceCube-Gen2 \cite{icecubegen2_whitepaper} will increase the number of astrophysical neutrinos with improved pointing information. This way, it is expected that neutrino astrophysics will contribute significantly to disentangling the contribution from Galactic and extragalactic sources in the energy region between the knee and the ankle.
\item The physics of {\bf charged cosmic rays} is proceeding rapidly, mainly through the measurement and modeling of the composition-dependent energy spectra of cosmic rays, see e.g.\ \cite{haungs2011,blasi2013,wolfendale2014} for reviews. Up to TeV-energies, direct measurements with balloons and satellites are possible. CREAM data reveal a break in the nuclei spectra for helium and heavier nuclei, which occurs around 100 GeV to 1 TeV in energy per nucleus \cite{ahn2010,biermann_apj2010}. Newer AMS02-data  provide an update of the proton\cite{ams02_protons_icrc2013} and helium \cite{ams02_helium_icrc2013} spectra at high precision, where the break seen in CREAM data could not be confirmed for helium yet. Indirect measurements of cosmic ray air showers at above TeV energies now also provide us with information about the light and heavy components of the cosmic ray spectrum. Using KASCADE and KASCADE-Grande, first evidence of the concrete the composition around the cosmic ray knee and above could be reconstructed \cite{kascade_grande2013,kascade_grande2014}. These air-shower data reveal a possible iron-knee at close to $10^{17}$~eV, indicating that the composition above the knee is actually becoming heavy as it would be expected by a spectral cutoff proportional to the charge of the nucleus.  However, the exact position of this iron knee, the question of the spectral behavior and composition at higher energies and other details still have to be resolved. Future IceTop data, including information on the composition, will help to answer this question \cite{aartsen_icetop2013}. As the region between the knee and the ankle is expected to be the transition region between Galactic and extragalactic cosmic rays, it is important to analyze the spectral behavior and composition above the ankle as well in order to get a handle on the extragalactic contribution. Recent results by the Auger collaboration show that even here, the composition seems to become heavier towards the high-energy cutoff of the cosmic ray spectrum \cite{auger_composition_icrc2013}.
\end{enumerate}
Composition-dependent cosmic ray transport through the Galaxy has been studied intensively in the past years, leading to the development of different numerical tools \cite{strong_propagation_1998,dragon,kissmann2014}, for a detailed description, see Section \ref{method:sec}. The GALPROP code was developed in the 1990s and for the first time provided a tool with realistic molecular distributions in the Galaxy, necessary to describe among others the longitude- and latitude-dependent, diffuse gamma-ray emission in the GeV energy range \cite{strong_propagation_1998}.
The transport equation solved numerically in GALPROP has the options to include a source term from
discrete sources and/or a source distribution, diffusion, convection,
diffusive re-acceleration, energy loss, fragmentation and radioactive
decay. The GALPROP code includes propagation in 3D, as well as
3D-maps of the Galactic magnetic field and the gas in the Galaxy. It
also includes the interstellar radiation field, which contributes to
inverse Compton scattering. 

GALPROP has among others
been used to estimate the production of cosmic ray secondaries like
gamma-rays from $\pi^0-$decays and positrons from charged pion
decays. In particular, it can be used to fit the observed gamma-ray
sky, see e.g.\ \cite{strong_new_1998,abdo_milagro2007,abdo_milagro2008}. Further, the code is
applied to determine the extragalactic gamma-ray background
\citep{abdo_extragalactic2010}. It has also been shown that, for a given
  source distribution function using a diffusive transport equation, 
the Milky Way behaves like a
  calorimeter \citep{strong2010}. In the modeling, a
  generic cosmic ray spectrum and with no discrete, known sources is typically assumed. Approaches
  using discrete, nearby supernova remnants with a generic
  cosmic ray spectrum with a diffusive transport equation were used in \cite{buesching2008} and \cite{ahlers2009} and could
  reproduce the observed increasing ratio of positrons to electrons and positrons. Similar results are
  found in calculations for SNRs from heavy stars 
  \citep{erlykin2013,biermann_prl2009}.
The PAMELA data show that
  the contribution of positrons to the electron and positron spectrum
  is rising towards higher energies \citep{pamela2009}, which cannot be explained by the
  typical results received with the GALPROP code, using a distribution
  of sources and without including discrete sources in the
  neighborhood. Results on the electron and positron spectra are now available form the AMS-02 experiments, improving the results and confirming the unexpected trend of an increasing positron fraction, which does not show a clear sign of a turnover yet \cite{ams_positrons2013}. 
Including positrons produced in cosmic ray
  interactions in nearby SNRs can produce the signature. Further
  observation features like the excess of microwave- and
  gamma-rays emission in the Galactic halo, named the Fermi-bubbles
  (GeV energies) and the WMAP-haze ($22-90$~GHz), could be due
  to an enhanced number of SNRs towards the Galactic center
  \citep{biermann_apjl2010}. In \cite{finkbeiner_galactic_wind2010}, the signal
  is interpreted as a possible signature from AGN activity or from a
  bipolar Galactic wind. 
A similar deviation from expected cosmic ray features is observed in
the arrival direction of cosmic rays. Several experiments observe a large-scale anisotropy in the cosmic ray
spectrum at around $10$~TeV, which is at a level of $\sim 6\cdot
10^{-4}$ \citep[e.g.]{amenomori_ta_anisotropy2006,guillian_superk_anisotropy2007,abdo_milagro_anisotropy2009,abbasi2010}. The feature cannot
be explained by the Compton-Getting effect, i.e.\ the motion of the
solar system through the Galaxy, which implies that the feature must
arise from the interstellar medium. Either the magnetic field
structure or the source distribution can be the reason \cite{giacinti_sigl2012,sveshnikova2013,pohl2013}. Alternatively, a local stream from an ancient supernova remnant in our local environment can explain the anisotropy \cite{biermann_anisotropie}.

In the recent years, in a huge effort to improve numerical modeling of cosmic ray transport, two other tools have been developed, both providing a cross-check for the GALPROP results 
and also providing other, improved features. The DRAGON code builds on previous GALPROP results and includes major improvements including a generalization of the diffusion process by introducing a radial dependence of the diffusion
coefficient \cite{dragon,gaggero2013}. The PICARD code was developed in order to be able to use a full diffusion tensor \cite{kissmann2014}. Here, a large effort was put on the implementation modern numerical methods to solve the partial differential equation and by that improving performance \cite{kissmann2014}. Both codes have included a spiral structure of the Milky Way \cite{gaggero2014,werner2015}. This way, electron and positron spectra including the new features revealed by AMS02 and PAMELA can be described with more realistic primary spectra as before.

In previous investigations with the different propagation tools, the normalization of the hadronic cosmic ray spectrum in the numerical framework is typically done using the total observed cosmic ray energy at Earth. No information from individual sources is considered and the normalization to the spectrum itself has therefore not been a major focus of previous work. In this paper, we focus on investigating the spectral behavior and the total cosmic ray energy budget to determine if observations of these two quantities match the hypothesis that supernova remnants are the sources of cosmic rays. 

While there is no unambiguous proof yet, SNRs are the most promising source class to explain the cosmic ray energy budget below the knee. In a simplified calculation, it is assumed that a typical supernova explosion provides a kinetic energy budget of $E_{\rm SN}\sim 10^{51}$~erg. If the cosmic ray spectrum below the knee represents a Galactic cosmic ray flux focussed within the Galactic plane, the inferred cosmic ray luminosity in the Galaxy is approximately $L_{\rm CR}\sim 2\cdot  10^{41}$~erg/s, within an uncertainty of about an order of magnitude as derived in e.g.\ \cite{drury2014}. Supernova explosions occur at an approximate rate of $\dot{n}\sim (1/50-1/100)$yr$^{-1}$ \footnote{In \cite{diehl2006}, the core collapse supernova rate is derived to be $1.9\pm1.1$ per century. We assume approximately $1-2$ SN per century which is compatible within the uncertainties.}. Assuming now that a constant fraction $\eta$ of the kinetic energy is converted to hadronic cosmic rays, it can be shown that $\eta$ needs to be on the order of $10\%$ in order for SNRs to explain the total cosmic ray energy budget,
\begin{equation}
  L_{\rm CR}\approx 2\cdot 10^{41}\,{\rm erg/s}\cdot \left(\frac{\eta}{0.1}\right)\cdot \left(\frac{\dot{n}}{0.02\,{\rm yr}^{-1}}\right)\cdot \left(\frac{E_{\rm SN}}{10^{51}\,{\rm erg}}\right)\,.
  \label{cr_lumi:equ}
\end{equation}

While the above presented calculation shows qualitatively that the total cosmic ray energy budget up to the knee can be produced by SNRs, a quantitative proof using realistic SNR energy spectra has not been possible. In particular, this back-of-the-envelope calculation assumes that all SNRs have (a) the same energy budget; (b) the same spectral index; (c) the same maximum energy. The theory of particle acceleration in SNRs, however, suggests that the SNR cosmic ray spectra actually change with time, both concerning all three parameters. A mixture of these parameters will therefore provide the average propagated cosmic ray energy spectrum. It is important to show that a realistic distribution of parameters actually does lead to the correct spectral behavior and energy budget.

In this paper, we use the proton spectra derived gamma-ray data from 21 well-studied SNRs in \cite{mandelartz2015} in order to investigate if the class of gamma-ray emitting SNRs can account for the cosmic ray proton energy budget. It is the first time that pieces of information from such a large sample of individual remnants is available. The studied population includes about $10\%$ of all supernova remnants which should be active simultaneously in the Galaxy\footnote{Here, we assume a rate of supernova explosions of $\dot{n}\sim 0.01-0.02$~yr$^{-1}$ with effective acceleration in the Sedov-Taylor phase over approximately $\tau_{\rm SN}\sim 10,000$~years. A total number of approximately $N_{\rm SNR}=100-200$ SNRs should therefore be active cosmic ray accelerators at a given time}.  The observed cosmic ray budget comes from the diffused pool of cosmic rays in the Galaxy, produced within the lifetime of cosmic rays in the Galaxy, i.e.\ $\tau_{\rm esc} \sim 10^{7}$~years \cite{yanasak2001}. As an individual SNR is only active for $\tau_{\rm SN}\sim 10^{4}$~years, the observed cosmic ray flux will be the average flux of a mixture of hundreds of sets of SNRs as we see them today, but with spectra and energy budgets distributed differently as they are today. In this approach, we assume that the $\gamma-$ray detected sample is representative for the entire population of SNRs at a fixed time. We therefore use the $N_{\gamma\rm{-SNR}}=21$ spectra and simulate the propagation of the local spectra through the Galaxy $m$ times (with $m$ as a large number, as described later in the paper). Finally, we reweight the spectrum by dividing by the number of SNRs simulated, $m$, and multiplying by the actual number of SNRs that are expected to be active at a time, i.e.\ $N_{\rm SNR} = 100$. This procedure will be described in more detail later when describing the method, in Section \ref{method:sec}. In following this procedure, the concrete SNRs are placed randomly in the Galaxy, with a weight corresponding to the expected supernova explosion density in the Galaxy. The resulting cosmic ray spectrum is given by the average flux of these populations. For our simulation, we use the GALPROP code \cite{Galprop_Web_Standford,strong_propagation_1998,strong_new_1998}. This way, we can use the spectra observed at this instant of time and derive an average spectrum produced on average within the lifetime of cosmic rays.

As a result from these simulations, we draw conclusions if and under what constraints the observed population of gamma-ray emitting SNRs can represent the population responsible for the cosmic ray flux below the knee. With our simulations, we can both test the spectral behavior and the total energy budget of the sources.

This paper is organized as follows: In Section \ref{method:sec}, we describe the approach used to achieve our goal, including a short description of the tool GALPROP for Galactic cosmic ray propagation and a detailed description of the implementation of the SNR spectra with individual spectral normalizations. In the same section, the back of the envelop calculation is tested in order to validate the method. The results of our simulations using the individual SNRs are presented in Section \ref{results:sec} and they are discussed in detail in Section \ref{discussion:sec}.

\section{Method \label{method:sec}}

\subsection{Description of the numerical approach }
In order to estimate the gamma-ray emitting SNRs' contribution to the observed cosmic ray flux, we use the GALPROP tool \cite{Galprop_Web_Standford,strong_propagation_1998,strong_new_1998} with its previous applications summarized in Section \ref{intro:sec}. As our results will mostly concern the discussion of the energy behavior of a constant diffusion coefficient in order to keep the number of free parameters to a minimum, we will use the GALPROP software in this paper. Additional features as provided by DRAGON and PICARD are not considered at this stage and might become interesting to take into account in future investigations. In the GALPROP program,
the transport equation is solved numerically using the Crank-Nicholson method, including the following terms:

\begin{eqnarray}
\frac{dn}{dt}(\vec{r},t,E)&=&Q(\vec{r},t,E)+\nabla
(D_{xx}\nabla{n}-\vec{U}\,n))+\frac{\partial}{\partial
  p}\left[p^{2}\,D_{pp}\, \frac{\partial}{\partial
    p}\frac{n}{p^2}\right]\nonumber\\
&-&\frac{\partial}{\partial
  p}\left((\frac{dp}{dt}-\frac{p}{3}\nabla\cdot\vec{U})\cdot n\right)-\frac{n}{\tau_f}-\frac{n}{\tau_d}\,.
\end{eqnarray}
Here, $p$ is the momentum of the particle, $n$ is the particle
density per momentum at a given point in space $r$ and $Q$ is a cosmic ray source spectrum at the source. The latter can either be represented by discrete sources or a continuous source distribution. Plain, scalar diffusion and diffusive re-acceleration is modeled with the constant
coefficients $D_{xx}$ and $D_{pp}$, respectively. The coefficients are usually modeled to match the observed secondary-to-primary ratio of cosmic rays \citep{strong_propagation_1998} and we will discuss details of what exact parameters we use in this simulation below. The velocity
$\vec{U}$ is the drift velocity of the particles in case of convection. Fragmentation and
radioactive decay happen on time scales of $\tau_f$ and $\tau_d$,
respectively. 
Focused acceleration as discussed in \cite{schlickeiser_jenko2010} is not considered here. Particle species considered are leptons and hadrons, and
their secondaries through propagation. All parameters depend on the particle species under
consideration. The
output of the program includes hadronic and leptonic spectra, as well as
the gamma-ray emissivity in every grid point. The latter results from
synchrotron radiation, bremsstrahlung, inverse Compton scattering and
hadronic interactions. All details of the program can be found at \cite{Galprop_Web_Standford,strong_propagation_1998,strong_new_1998}. The specific settings and concrete method applied in this paper is described below.

\subsubsection{Cosmic ray spectra from gamma-ray measurements}

In this paper, gamma-ray observations are used to derive proton spectra from individual SNRs, observed in the Milky Way at this instant of time. We use the proton spectra for 21 SNRs as derived in \cite{mandelartz2015}, assuming that the sample is representative for a SNR population at a given time. The source spectra $j_{p}(T)$, with $T$ as the kinetic energy of the particle, and with $[j_p]=1/$MeV are parametrized as follows:
\begin{equation}
j_{\rm p}(T)=a_{p}\ \sqrt{\frac{T^2+2Tmc^2}{T_{0}^2+2T_{0}mc^2}}\ \frac{T+2Tmc^2}{\sqrt{T^2+2Tmc^2}}\ \tanh\left( \frac{T}{T_{\min}} \right)\ \exp\left(-\frac{T}{T_{\max}} \right).
\label{eq:SourceSpec_Mandelartz}
\end{equation}
Here, $a_p$ and $m$ represent the normalization constant and the proton mass, respectively. In \cite{mandelartz2015}, a low energy cutoff is applied at $T_{\min}=10$~MeV via a tangent hyperbolicus function in order to have a smooth transition. At high energies, the cutoff at kinetic energy $T_{\max}$ is performed via an exponential function. As we focus on the CR energy range from $\sim$GeV to $\sim$PeV in this paper, the low energy cut off is not applied in our simulations. The reference energy $T_0$ appears for a simpler parametrization of the function and is chosen to be $T_0=1$~TeV. Table \ref{params_source_spectra:tab} summarized the basic input parameters for the individual source spectra as they are provided by \cite{mandelartz2015}.

\begin{figure}
  \begin{center}
    \includegraphics[width=0.8\textwidth]{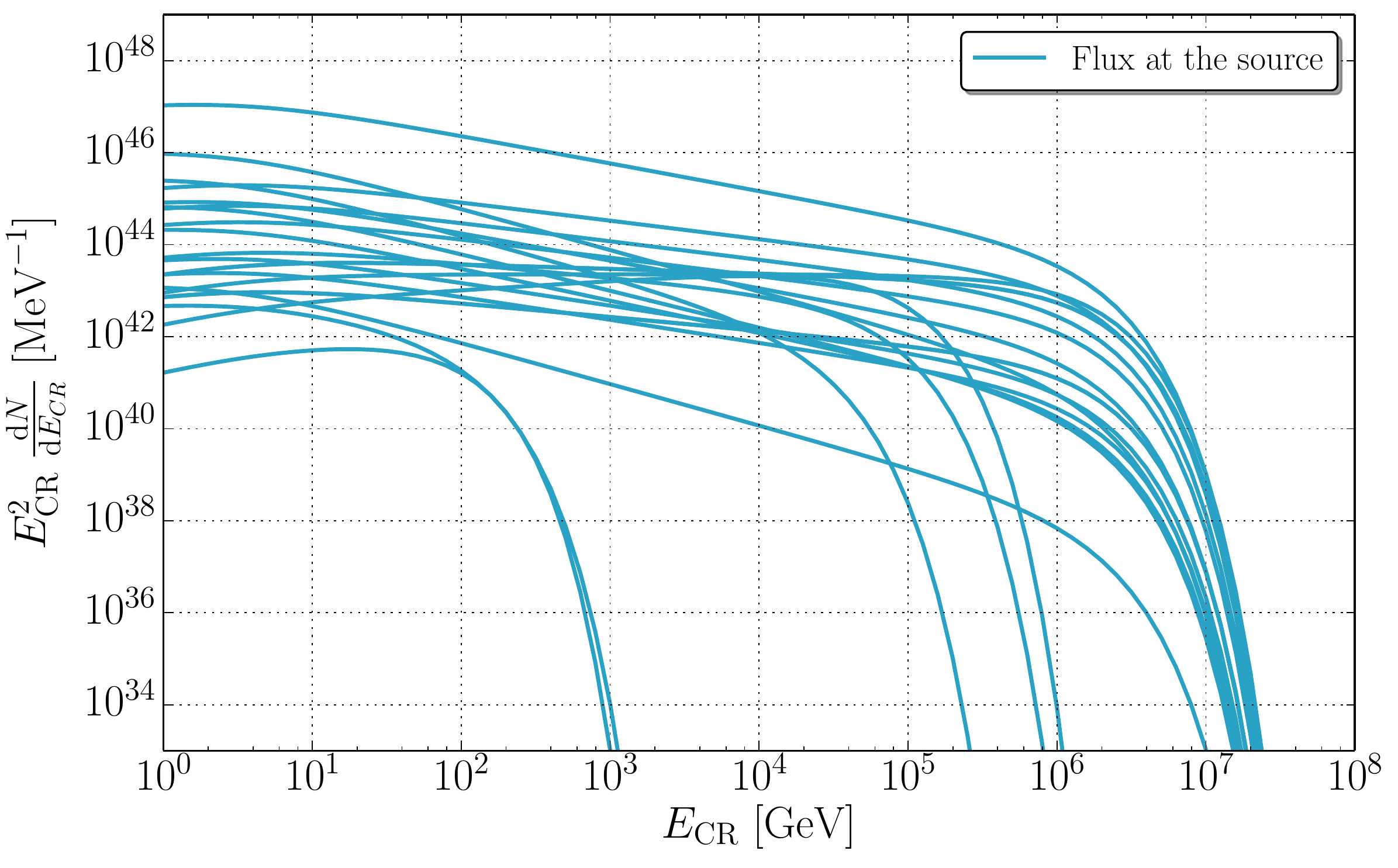}
  \end{center}
  \caption{\label{snrs:fig}Figure of SNR proton spectra {\it at the source}. These spectra are used to be propagated within GALPROP and to estimate the contribution of SNRs to the detected CR spectrum.}
\end{figure}

\begin{table}
\centering{
\begin{tabular}{c|cccccccc}
\hline\hline
SNR&$d$&$t_{\rm age}$&$\alpha_p$&$a_p$&$T_{\rm max}$&$E_{\rm CR,tot}$&RA&Dec\\
&[kpc]&[kyr]&&[$10^{39}$/MeV]&[GeV]&$10^{47}$~erg&&\\\hline
3C391 & 7.2 & 4.0 & 2.6&44964.2&$10^{6}$ & 3081.2&18h 49m 25s & -00$^{\circ}$ 55' 00" \\
W41 & 4.2 & 100.0 & 2.4&52175.2&$10^{6}$ & 4438.1&18h 34m 45s & -08$^{\circ}$ 48' 00" \\
W33 & 4.0 & 1.2 & 2.1&29694.1&$10^{6}$ &966.0&18h 13m 37s & -17$^{\circ}$ 49' 00" \\
W30 & 4.0 & 25.0 & 2.9&19853.4&$1.4\cdot 10^{4}$ &681.9& 18h 05m 30s & -21$^{\circ}$ 26' 00" \\
W28 & 1.9 & 33.0 & 2.8&9952.4&$10^{6}$ & 1874.6&18h 00m 30s & -23$^{\circ}$ 26' 00" \\
W28C & 1.9 & n/a& 2.5&2331.8&$10^{6}$ & 29.3&17h 58m 56s & -24$^{\circ}$ 03' 49" \\
G349.7+0.2 & 18.3 & 10.0 & 2.4 &332128.6& $10^{6}$&3155.2 & 17h 17m 59s & -37$^{\circ}$ 26' 00" \\
CTB 37B & 13.2 & 1.8 & 2.1&29721.8&$10^{6}$ & 3745.9&17h 13m 55s & -38$^{\circ}$ 11' 00" \\
CTB 37A & 7.9 & 16.0 & 2.6&92854&$10^{6}$ &1241.3& 17h 14m 06s & -38$^{\circ}$ 32' 00" \\
SN 1006 & 2.2 & 1.0 & 2.3&2676.1&$10^{6}$&1227.6& 15h 02m 50s & -41$^{\circ}$ 56' 00" \\
Puppis A & 2.0 & 4.6 & 2.5&4719.8&$10^{6}$ &231.2& 08h 22m 10s & -43$^{\circ}$ 00' 00" \\
Vela Jr & 1.3 & 4.8 & 1.8&16348.6&$4.4\cdot 10^{4}$ &1389.6& 08h 52m 00s & -46$^{\circ}$ 20' 00" \\
MSH 11-62 & 6.2 & 1.3 & 1.7&2869.8&46.0 &4.2& 11h 11m 54s & -60$^{\circ}$ 38' 00" \\
W44 & 3.0 & 10.0 & 2.6&258.4&58.7 &1.1& 18h 56m 00s & 01$^{\circ}$ 22' 00" \\
G40.5-0.5 & 3.4 & 30.0 &2.0&22697.4&$10^{6}$ &71.2& 19h 07m 10s & 06$^{\circ}$ 31' 00" \\
W49B & 10.0 & 1.0 & 2.9&76237.4&$10^{6}$&1323.3& 19h 11m 08s & 09$^{\circ}$ 06' 00" \\
W51C & 6.0 & 26.0 & 2.4&118406.8&$10^{6}$ &7872.5& 19h 23m 50s & 14$^{\circ}$ 06' 00"\\
IC443 & 1.5 & 3.0 & 2.7&6046.8&$10^{6}$ &85.2& 06h 17m 00s & 22$^{\circ}$ 34' 00" \\
Cygnus Loop & 0.6 & 15.0 & 2.9&93.2&$10^{6}$ &251.9& 20h 51m 00s & 30$^{\circ}$ 40' 00" \\
Cas A & 3.5 & 0.3 & 2.3&19276.6&$3.7\cdot 10^{4}$&2317.8& 23h 23m 26s & 58$^{\circ}$ 48' 00" \\
Tycho & 3.5 & 0.4 & 2.3&2678.0&$10^{6}$&1813.6& 00h 25m 18s & 64$^{\circ}$ 09' 00" \\
\end{tabular}
\caption{Basic input parameters of the 21 remnants in used here: $d$ is the distance to the SNR, $t_{\rm age}$ gives the SNR's age, $\alpha_p$ is the spectral index of the source spectrum, while $a_p$ gives the normalization constant at the reference kinetic energy $T_0=1$~TeV. Further, $T_{\rm max}$ represents the maximum energy of the spectrum, which is taken to be $1$~PeV in those cases where no clear cutoff could be identified in the data. $E_{\rm CR, tot}=\eta\cdot E_{\rm SNR}$ represents the total energy budget going into cosmic rays. RA/Dec provide right ascension and declination. All parameters are taken from the work of \cite{mandelartz2015}.\label{params_source_spectra:tab}}
}
\end{table}
Figure \ref{snrs:fig} shows the proton spectra {\it at the source}. The individual spectra differ significantly from each other: Their total cosmic ray energy budget varies from $10^{47}$~erg to $>10^{50}$~erg. Some spectra show a clear, early cutoff, others are flat and allow for a spectrum that continues up to the cosmic ray knee, i.e.\ up to $10^{15}$~eV in energy. For those sources that do not reveal a cutoff in gamma-ray data at this point, we assume that they continue up to the knee. This is the most optimistic scenario. It can be expected that in reality, a fraction of these sources actually has lower maximum energies, but at this point, it cannot be derived which ones and how many. This is why we consider this paper as a first, maximum scenario of how much these sources can contribute to the CR spectrum. In the future, when data from HAWC and CTA are available, this analysis can be redone with higher precision.

\subsubsection{Simulation of individual SNRs as sources of the cosmic ray flux at Earth}

The cosmic ray spectrum at Earth represents the average of cosmic ray spectra, over the time period that cosmic rays diffuse through the Galaxy, i.e.\ $\tau_{\rm esc}\approx 10^{7}$~years. During this time, $\sim 100$ SNR populations contribute to the average cosmic ray flux. The positions of these SNRs follow the distribution of massive stars in the Galaxy, assuming that the SNRs are active for $10,000$~years or more. Here, we describe the procedure of how we use the sample of today's population to calculate the diffuse cosmic ray flux that has been averaged from earlier SNRs over the years:

\begin{enumerate}
\item We use those $N_{\gamma-{\rm SNR}} =21$ spectra derived in \cite{mandelartz2015} as representative for one SNR population. The sample discussed above only makes up  a fraction of the total population, as the sensitivity of gamma-ray telescopes is basically limited to some $8-10$~kpc distance from Earth and does not provide data for typical SNRs from across the Galaxy. If we assume that the supernova rate in the Galaxy is $\dot{n}\approx 0.01-0.02$~yr$^{-1}$, and that one SNR actively accelerated cosmic rays in the Sedov-Taylor phase, lasting $t_{\rm SNR}\approx 10,000$~yrs, there are about $N_{\rm SNR}\approx 100-200$~active cosmic ray emitters at a given time. This is consistent with the number of shell-type SNRs that are well-identified at radio energies and catalogized in the Green catalog \cite{green2014}. Here, the number is close to 300. However, it should be kept in mind that radio emission is expected to happen even after the Sedov-Taylor phase, up to $10^{5}$ years or longer. So, while a larger number of radio SNRs than documented in Green's catalog is actually expected to be present in the Galaxy, only the stronger radio emitters are expected to be cosmic ray emitters and the detected numbers of strong radio emitters therefore provide a rough cross-check of active cosmic ray emitters. 

The sample of SNRs we use here therefore provides us with a fraction 
\begin{equation}
1/\alpha=:\frac{N_{\gamma-{\rm SNR}}}{N_{\rm SNR}}\approx 1/8
\end{equation}
of all SNRs that contribute to the cosmic ray spectrum.
Once we receive or final result of the cosmic ray proton spectrum at Earth from those 21 SNRs, we therefore have to weight the normalization with a factor of  $\alpha$ in order to calculate the total energy output of all SNRs in the Galaxy.
\item In order to take into account the fact that the cosmic ray spectrum observed at Earth today represents the average of SNRs active for the past $10^{7}$~years, we place those spectra with individual normalization and spectral index at random positions in the Galaxy. It is not known, where in the past $10^{7}$~years SNRs have been active, but the distribution of supernova explosions should follow the mass distribution in the Galaxy. We therefore use this distribution in order to weight the probability of an SNRs being placed on the grid of the Galaxy in the GALPROP simulation. Specifically, we use the distribution function of  \cite{1998ApJ...504..761C} implemented in GALPROP to implement the weighting. It is clear at this point, that this description is not optimal and does not fully represent the true distribution of SNRs \cite{2009BASI...37...45G}. At this point, it will be considered as a first-order approximation. Future work will include more precise distributions as for instance presented in \cite{2009BASI...37...45G}. For the first order approach that we follow here, the precision of the distribution presented in \cite{2009BASI...37...45G} suffices. As this provides us with a number in $(x,y)$ out of $\mathbb{R}^2$, but GALPROP as a numerical tool parametrizes the Galaxy on a grid with a certain spacing $(\delta x,\,\Delta y)$, we move the source simulated at a random position with a distribution-weight to the grid-point closest to the result. As the grid-size is typically significantly larger than the size of a single SNR, i.e. $\Delta x,\,\Delta y \gg 10$~pc, this implies we simulate SNRs as point-like sources.
\item
  For statistical reasons we simulate a large number of supernova remnants, $m$, that blow up simultaneously and are active for 10,000 years. Each of the SNRs randomly receives one of the energy spectra drawn from the set of 21 SNRs. The cosmic ray flux generated by these $m$ SNRs will then be weighed by the average number of SNRs considered to be active in a certain time frame: As a total number of SNR we choose $m = 10,000$ or $m = 20,000$ (depending on the specific simulation)  and we will show in the result section that this provides us with reasonable statistics. For the true, average number of SNRs active in the Galaxy at one time, we use $N_{\mathrm{SNR}} = 100$. As explained earlier, the detected number of SNRs in the radio is close to 300, but it is expected that only the brightest radio sources actually do contribute significantly to the total flux of cosmic rays, as these represent the younger SNRs that are able to accelerate to high energies. There is obviously an uncertainty attached to this number, which we consider to be a factor of $\sim 2$. Hence, in order to have a correct scaling of the normalization, the resulting properly normalized cosmic ray flux is obtained by reweighting the flux that is simulated with $m$ SNRs, $\Phi_{m}$ by the fraction $N_{\rm SNR}/m$, e.g.\ as
  \begin{equation}
    \Phi_{\mathrm{res}} = \frac{N_{\rm SNR}}{m}\times\Phi_{m} = \frac{100}{10,000}\times\Phi_{m}\,.
\label{phi:equ}
  \end{equation}

\end{enumerate}

This procedure is straight forward, however, a few things should be kept in mind: The sample of 21 SNRs used to derive individual CR spectra at this point still only represents about 1/5th to 1/10th of the total SNR population. It is not entirely clear yet if this sample is fully representative for the population of cosmic ray emitting SNRs. Gamma-ray observations are done with good energy and spatial precision in the GeV (Fermi) and TeV (H.E.S.S./MAGIC/VERITAS) ranges. This corresponds to cosmic ray energies from around a few GeV up to approximately 100 TeV and thus does not include the knee region. There may exist flat cosmic ray sources that are not very prominent at below TeV gamma-ray energies, but which, due to their flat spectrum, contribute to the total cosmic ray spectrum. Thus, today's sample may not be statistically complete. Future measurements with HAWC \cite{hawc} and CTA \cite{cta} will help to provide a complete sample, as with these next generation telescopes, the sensitivity will be enhanced to have a full view of the entire Galaxy in gamma-rays, and the energy range is expected to be increased up to values of  $100-300$~TeV. The analysis of the spectrum done here should therefore be considered as a starting point - in a few years from now, this procedure can be repeated, with higher statistical significance and more knowledge about SNRs as potential sources of Galactic CRs.

\subsubsection{Normalization of individual SNR spectra in GALPROP}
First of all it should be noted that the CR normalization in \textit{standard} GALPROP applications is usually fixed by globally scaling the calculated cosmic ray density such that the observed CR flux at Earth is met. In the approach presented in this paper, the idea is to fix the normalization by the rate of particle injection of the individual SNRs into the Galaxy.
Hence, from the technical viewpoint a key ingredient is to rewrite the source spectrum  $j_{\rm p}(T)$ in terms of the \textit{internal} GALPROP units. A pragmatic approach to determine the needed conversion factor is to compare the calculation of the luminosity $L$ in GALPROP with the corresponding integral expression on the basis of $j_{\rm p}(T)$
\begin{equation}
L = \xi R^2_{\rm SNR}c \int_{10~\rm{MeV}}^{1e9~\rm{MeV}}\ dT\ \beta\ T\ j_{\rm p}(T)/V_{\rm SNR}\,. 
\label{eq:LuminosityGeneral1}
\end{equation}   
Here, $V_{\rm SNR}=4/3\,\pi\,R_{\rm SNR}^{3}$ is the volume of the SNRs with a radius $R_{\rm SNR}$. Note that $\xi =1/2$ corresponds to the case where the CR particles stream out of the SNR in radial direction. In contrast to that $\xi = 1$ arises from an averaging assuming that the CRs move in random directions inside the SNRs. The aforementioned comparison leads to the following relation between the initial source function $q_1(p(T))$ as implemented in GALPROP. Further details can be found in the GALPROP documentation~\cite{strong_galprop_2011,Galprop_Web_Standford} or by directly checking the source code file cr\_luminosity.cc.

In this case, the spectrum parametrization  $j_{\rm p}(T)$ in GALPROP becomes
\begin{equation}
q_1(p(T))= \alpha \frac{c^{2} R_{\rm SNR}^{2} \beta^{2}}{4\pi V_{\rm grid}}\ j_{\rm p}(T).
\label{eq:ConversionISF}
\end{equation}

In an alternative approach, the luminosity can be derived from the total energy $E_{\rm tot}$ of protons in the SNR
via $L=E_{\rm tot}/\tau$ with a time-scale $\tau$, representing the distribution of the total energy over the total  lifetime of the remnant,
\begin{equation}
L =  \frac{1}{\tau} \int_{10~\rm{MeV}}^{1e9~\rm{MeV}}\ dT\ T\ j_{\rm p}(T). 
\label{eq:LuminosityGeneral2}
\end{equation}
Using this expression for the luminosity, one finds
\begin{equation}
q'_1(p(T))= \frac{\beta c R_{\rm SNR}^{3}}{3 V_{\rm grid} \tau}\ j_{\rm p}(T).
\label{eq:ConversionISF_alter}
\end{equation}
Assuming that $E_{\rm tot}$ is the energy of the SNR converted into protons and $\tau$ is the life time of the SNR, $L=E_{\rm tot}/\tau$ can be interpreted as the average luminosity in CRs.

It is a typical approach in gamma-ray astronomy to derive the total cosmic ray energy budget in order to estimate the SNRs possible contribution to the total cosmic ray budget. A back-of the envelope calculation predicts that the observed cosmic ray luminosity of the Galaxy (about $3\cdot 10^{40}$~erg/s) can be reproduced if on average, $10^{50}$~erg are going into a single SNR at a Supernova rate of $(1/50-1/100)$~yr$^{-1}$. Here, it is assumed that cosmic rays are injected into the interaction region continuously at a constant rate during the lifetime $\tau$ with the total energy going into cosmic rays conserved over time. It is clear that this is a simplifying assumption, as it is known that at least the energy spectrum is changing with time, in particular concerning the reduction of maximum energy, see e.g.\ \cite{cox1972}. For energy spectra steeper than $E^{-2}$, the total energy budget is dominated by the lower integration threshold, so effects from this temporal development should be relatively small. It is also specified in \cite{cox1972} that the total energy of the SNR is decreasing with time due to cooling effects. This would mean that, if we assume a constant fraction of the SNR energy going into cosmic rays at a given time, the actual average luminosity of cosmic rays would be underestimated in particular for old remnants. It is not clear, however, if the fraction of energy going into cosmic rays is constant over time or if it actually decreases with the available energy budget. Thus, we judge that in first order approximation, it seems reasonable to estimate the total energy of the remnant from the given value, keeping in mind the above discussed uncertainties. In this paper, this normalization scheme, i.e.\ following the total energy argument, will be followed. This scheme normalizes the individual SNRs with respect to each other as well as each SNR individually, so that the final result will be both a realistic {\it spectral energy behavior} as well as {\it normalization} of the spectrum.

\subsubsection{Including CR Nuclei}
Although the source spectra taken from~\cite{mandelartz2015} are only provided for CR protons, it is possible to include CR nuclei in GALPROP simulations (see chapter 5.5 in ~\cite{strong_galprop_2011,Galprop_Web_Standford}). To do so, the initial source function of nuclei $q_{A}(p_{A})$ with mass number $A$ and momentum $p_{A}$ is related to $q_{1}(p_{1})$ by the relative abundance $X$ according to
\begin{equation}
X=\frac{Aq_{A}(p_{A})}{q_{1}(p_{1})}.
\label{eq:RelAbundanceAndNuclei}
\end{equation}
In this context two remarks have to be made: 
\begin{enumerate}
\item Due to the high energy cut off in equation (\ref{eq:SourceSpec_Mandelartz}) $X$ is not independent of the energy in what follows. 
\item Including CR nuclei injection, the total energy in hadrons of the SNRs is artificially increased. The energy budget can approximately be derived following \cite{mandelartz2015} by down-scaling the proton normalization $a_{p}$ in equation (\ref{eq:SourceSpec_Mandelartz}) appropriately.
\end{enumerate}
Here, simulations are performed for all nuclei, but the resulting energy spectra are only discussed for protons. In the future, the  heavy nuclei spectra will be discussed as well in order to investigate other questions connected to cosmic ray observations, sources and transport.
 Photohadronic effects as well as photo spallation is typically negligible at the given length scales and electromagnetic fields in the Galaxy.

 \subsubsection{GALPROP settings}
 GALPROP provides the user with a large number of parameters that can be changed to follow the users needs. Most of these were left unchanged with respect to version 54.1.984. In this section, we summarize what has been changed and which parameters were varied. All changes to the galdef-file are summarized in table \ref{default}. The main changes are discussed below:
 \begin{enumerate}
   \item {\bf The normalization scheme}\\
In order to investigate normalization and spectral behavior, the original GALPROP code must be modified. In its current version we are able to provide individual SNRs with their own parameters as been measured or obtained by specific analyses, see \cite{mandelartz2015}. The sources are injected with their spatial parameters, such as their distance to Earth and their actual extension. Each SNR is provided with a set of spectrum normalization, spectral index and maximum energy, provided by 21 individual sets given in \cite{mandelartz2015}. Here, for each SNR, the position is injected randomly, weighted by a source distribution function. As GALPROP propagates particles on a grid, each randomly drawn position is internally set to the closest grid point.

\item {\bf The Galaxy size}\\
  The Galaxy is treated as a three dimensional object and the grid points are arranged with a distance of $d_{\mathrm grid} = 1$ kpc from each other. Thus, the SNRs can effectively be seen as point-like sources. In our analysis, we tested two Galaxy sizes. In the small galaxy configuration the horizontal plane ranges from $-10 \mathrm{\ kpc} \leq (x,y) \leq 10$ kpc, while the large Galaxy doubles the small Galaxy in each direction. We use the small Galaxy size only for first tests as simulations are quicker for this smaller size simulation. We show only the first graph of our results with both Galaxy sizes, as they deliver comparable results in all simulations. The vertical component remains constant for both configurations and extends over a range of  $-4 \mathrm{\ kpc} \leq z \leq 4$ kpc.

\item {\bf The diffusion coefficient}\\
  Furthermore, the diffusion coefficient has been varied in our simulations in order to analyze the energy budget of the resulting cosmic ray spectrum. We compare the Kolmorogov-type diffusion with $D_{xx} = E^{\delta}$ and $\delta = 0.33$ to a steeper diffusion with $\delta = 0.5$.\\

\item {\bf The individual supernova remnants}\\
  The analyses cover two settings of SNR types. The first can rather be seen as the standard SNR as it has a fixed total energy $E_{\rm CR, tot}=10^{50}$~erg and also a fixed spectral index, where three cases are investigated: $\alpha_p=2.0,\,2.3,\,2.5$. This is done in order to test if the back-of-the-envelop calculation presented in the introduction and often used as an argument that SNRs can be the sources of Galactic cosmic rays actually holds, even in this more advanced calculation.
  
The latter type takes into account the individual SNR parameters as derived in \cite{mandelartz2015}. It is the aim of this paper to investigate if these gamma-ray emitting SNRs can be the sources of Galactic cosmic rays. In order to quantify how reliable the results are with respect to the primary cosmic ray spectra as derived in \cite{mandelartz2015}, the primary cosmic ray spectral indices of all SNRs have been allowed to vary within a $1\sigma$ uncertainty. This way, for the final result, a 1$\sigma$ error band can be drawn and the uncertainty of the spectrum can be estimated.
 \end{enumerate}
\begin{longtable}{l|c|r|r}
\hline\hline
Parameter	&	Unit	&	Original Galdef	&	Our Modifications	\\
\hline
\endhead
\hline \multicolumn{4}{c}{\textit{Table continues on the next page}} \\
\endfoot
\endlastfoot
\multicolumn{3}{l}{Grid options and spectra}\\
\hline
r\_min	&	kpc	&	0.0	&	0.0	\\
r\_max	&	kpc	&	30.0	&	25.0	\\
x\_min	&	kpc	&	0.0	&	-20.0	\\
x\_max	&	kpc	&	+20.0	&	+20.0	\\
dx	&	kpc	&	0.2	&	1.0	\\
y\_min	&	kpc	&	0.0	&	-20.0	\\
y\_max	&	kpc	&	+20.0	&	+20.0	\\
dy	&	kpc	&	0.2	&	1.0	\\
z\_min	&	kpc	&	-4.0	&	-4.0	\\
z\_max 	&	kpc	&	+4.0	&	+4.0	\\
dz	&	kpc	&	0.1	&	0.2	\\
p\_min	&	MV	&	1000.0	&	1000.0	\\
p\_max	&	MV	&	4000.0	&	4000.0	\\
p\_factor	&		&	1.2	&	1.3	\\
Ekin\_min	&	MeV	&	1.0e1	&	1.0e1	\\
Ekin\_max	&	MeV	&	1.0e7	&	1.0e9	\\
Ekin\_Factor	&		&	1.2	&	1.3	\\
E\_gamma\_min	&	MeV	&	0.1	&	100.0	\\
E\_gamma\_max	&	MeV	&	1.0e6	&	1.0e6	\\
E\_gamma\_factor	&		&	10.0	&	1.5	\\
long\_min	&	deg	&	0.5	&	0.0	\\
long\_max	&	deg	&	359.5	&	360.0	\\
lat\_min	&	deg	&	-89.5	&	-90.0	\\
lat\_max	&	deg	&	+89.5	&	+90.0	\\
d\_long	&	deg	&	10.0	&	1.0	\\
d\_lat	&	deg	&	10.0	&	1.0	\\
healpix\_order	&		&	7.0	&	6.0	\\
\hline
\multicolumn{3}{l}{Cosmic Ray propagation parameters}	\\
\hline
nuc\_rigid\_br	&	MV	&	1.0e4	&	1.0e2	\\
nuc\_g\_1	&		&	2.23	&	2.43	\\
nuc\_g\_2	&		&	2.43	&	2.43	\\
inj\_spectrum\_type	&		&	rigidity	&	powerlaw	\\
electron\_g\_0	&		&	2.1	&	2.5	\\
electron\_rigid\_br0	&	MV	&	1.0e3	&	1.0e3	\\
electron\_g\_1	&		&	2.3	&	2.50	\\
electron\_rigid\_br	&	MV	&	1.0e4	&	1.0e3	\\
electron\_g\_2	&		&	2.50	&	2.50	\\
\hline
\multicolumn{3}{l}{Parameters controlling interstellar medium}	\\
\hline
He\_H\_ratio	&		&	0.11	&	0.11	\\
n\_X\_CO	&		&	10.0	&	9.0	\\
X\_CO	&		&	1.9e20	&	1.9e20	\\
max\_Z	&		&	28.0	&	1.0	\\
\hline
\multicolumn{3}{l}{Parameters controlling propagation calculation}	\\
\hline
start\_timestep	&	years	&	1.0e7	&	1.0e9	\\
end\_timestep	&	years	&	1.0e1	&	1.0e2	\\
timestep\_factor	&		&	0.5	&	0.5	\\
timestep\_repeat	&		&	20.0	&	20.0	\\
network\_iterations	&		&	1.0	&	2.0	\\
network\_iter\_compl	&		&	1.0	&	2.0	\\
\hline\hline
\caption{This table gives an shows parameters that have been changed due to the standard GALDEF file as can be found in the GALPROP manual \cite{Galprop_Web_Standford}.}
\label{default}
\end{longtable}%


\section{Results and Conclusions \label{results:sec}}
        
        In this section, we show the results from the simulation as described above. We focus on the energy range from 10 GeV upward, as gamma-ray measurements have precise results between $\approx 1$~GeV and $10$~TeV, corresponding to a cosmic ray energy of about $10$~GeV to $100$~TeV. Those spectra that do not show a cutoff up to $100$~TeV cosmic ray energy are extrapolated with an assumed cutoff in the knee region.
        
\subsection{Validation of the method \label{verification:sec}}
\begin{figure}
  \begin{center}
    \includegraphics[width=\textwidth]{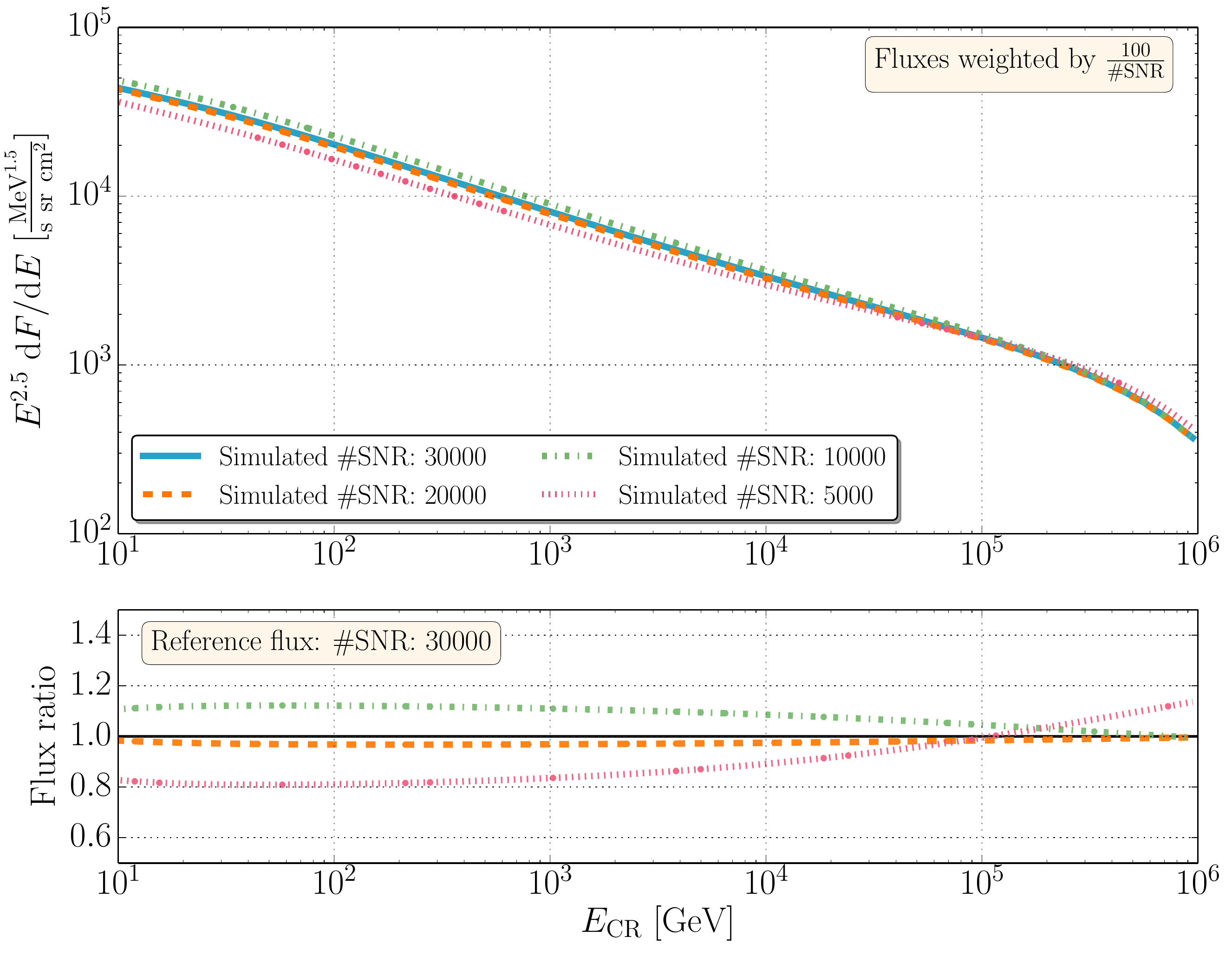}
  \end{center}
  \caption{\label{test:fig} {Simulation of individual SNRs, for $1,000,\,10,000,\,20,000$ and $30,000$ SNRs, always normalized to the true number, i.e.\ 100.}
    }
\end{figure}

In order to validate the approach chosen here, we use the full GALPROP simulation to test two things: first of all, we simulate individual spectra for $1,000,\,10,000,\,20,000$ and $30,000$  SNRs in order to cross-check the statistical convergence. Figure \ref{test:fig} shows the spectra for the different numbers (top) and the ratio with respect to the simulation of the highest number ($30,000$ SNRs, bottom). Here, we test how many SNRs have to be simulated in order to receive a statistically relevant result. The figure shows the CR spectrum for simulated $1,000,\,10,000,\,20,000$ and $30,000$ SNRs, always normalized to the true number of SNRs in the Galaxy at one time, see Equ.\ (\ref{phi:equ}). Thus, in the ideal case, they should give the same result, and higher numbers should result in a more precise calculation. This becomes evident when looking at the flux ratio, where deviations with respect to the largest number of simulated SNRs ($30,000$) become smaller. In the case of $20,000$ SNRs, deviations are on the order of 1\% and we use this number for all following calculations.

Secondly, we test the standard approach, i.e.\ all sources have the same spectral behavior, luminosity and maximum energy as it is presented in Section \ref{intro:sec}. Here, each simulated SNR receives the same spectrum, with a normalization corresponding to a total cosmic ray luminosity of $2\cdot 10^{41}$~erg/s (compare Equ.\ \ref{cr_lumi:equ}). Concretely, the spectral behavior is assumed to be a power-law with a spectral index $\gamma$ at the source and a maximum energy $E_{\max} = 10^{15}$~eV:
\begin{equation}
j_{\rm standard} = A\cdot \left(\frac{E}{E_0}\right)^{-\gamma}\cdot \exp\left(-\frac{E}{E_{\max}}\right)\,.
\end{equation}
    While the standard approach in text books does not rely on the spectral index, as it only concerns the total energy budget, we test three different indices at the sources, i.e.\ $\gamma=2.0,\, 2.3,\, 2.5$. The diffusion coefficient is assumed to follow an $E^{1/3}$-behavior, representing a Kolmogorov spectrum. In principle, the energy behavior could be as strong as $E^{0.6}$. But as the change in the primary cosmic ray spectrum towards steeper spectra has the same effect as changing the diffusion coefficient to a stronger energy behavior, we refrain from changing the diffusion coefficient in this test and only change the primary spectrum.

\begin{figure}
  \begin{center}
    \includegraphics[width=\textwidth]{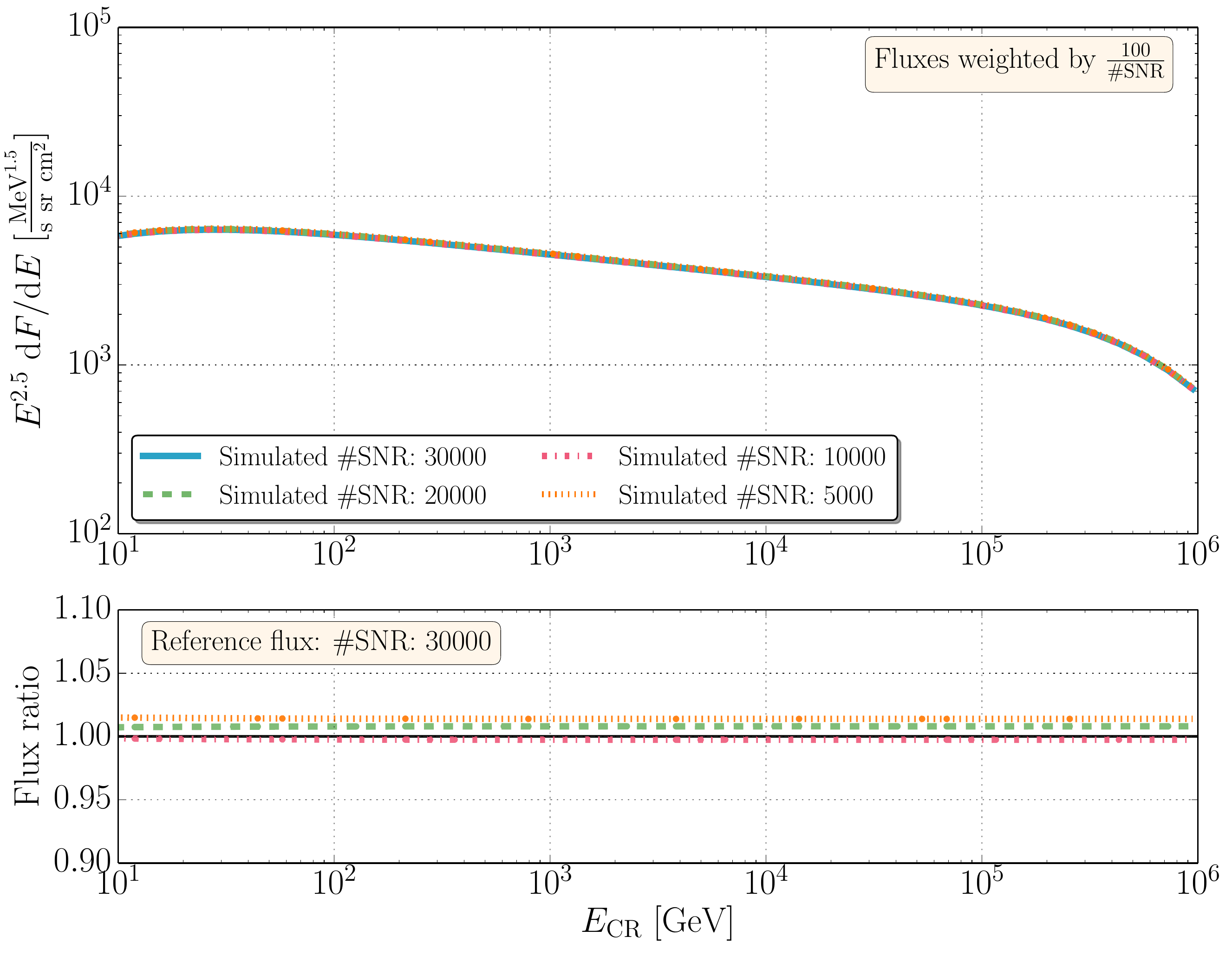}
  \end{center}
  \caption{\label{standard_no_snrs:fig} The standard approach with a generic $E^{-2.3}$ spectrum and the same normalization for every SNR.  Here, the total number of 10,000 simulated supernova events is sufficient as the statistical error reduces to below $1\%$ between a simulation of $10,000$ and $30,000$ SNRs.}
\end{figure}

Figure \ref{standard_no_snrs:fig} shows how the spectra change for different number of simulated SNRs, i.e.\  $1,000,\,10,000,\,20,000$ and $30,000$ for the case of $E^{-2.3}$. It can be shown that the spectra converge for larger numbers of SNRs and that the differences between the simulation of $10,000$ and $20,000$ SNRs are below 1\%. We therefore use $10,000$ SNRs for this simulation in the following.

\begin{figure}
  \begin{center}
    \includegraphics[width=\textwidth]{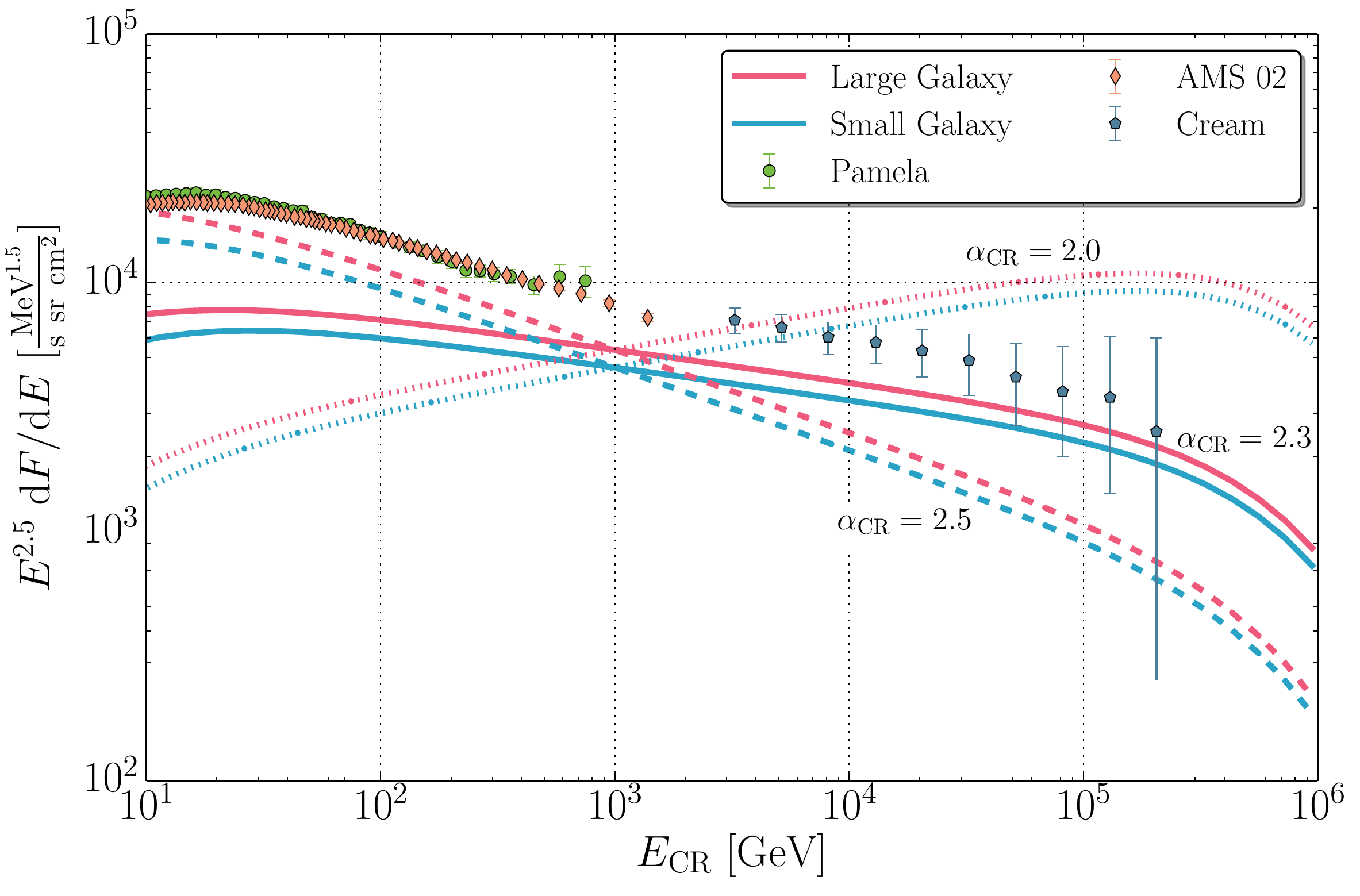}
  \end{center}
  \caption{\label{standard:fig} {This figure shows the comparison of the cosmic ray flux between different generic spectral indices, $E^{-2.0}$, $E^{-2.3}$ and $E^{-2.5}$ by using the standard approach. The blue lines show the small galaxy, while the blue line represents the large galaxy configuration.}
    }
\end{figure}

    Figure \ref{standard:fig} shows the spectra for the large (upper line in each set) and small (lower line) Galaxy in the standard approach, where all sources are treated the same. Firstly, it can be noted that the differences of the spectra for the small and the large Galaxy are rather small and can be considered as negligible given the larger uncertainties of the input parameters. We therefore only show results for the large Galaxy in the following sections. Secondly, it is shown here that the cosmic ray flux is generally underestimated by a factor of a few (approximately $2-5$). Only for very flat injection spectra, i.e.\ $E^{-2}$, in combination with Kolmogorov-type diffusion results in an overestimation at the highest energies. This combination is, however, the flattest spectrum that is generally possible and it actually underestimates the energy budget at the lowest energies by a large amount. This is therefore not a realistic scenario. For steeper spectra (or larger energy dependence in diffusion), a deviation from data by a factor of $2-4$ is certainly within the errors of this calculation: parameters like the supernova rate and the average energy budget of the SNRs are highly uncertain. Thus, we consider this result as compatible with the back-of-the-envelop calculation concerning the energy budget. The spectral behavior, on the other hand, does not follow a single power-law. For low energies, i.e.\ below a few TeV it matches an $E^{-2.5}$ primary spectrum (solid line). At higher energies, the spectrum is better represented by an $E^{-2.3}$ spectrum (dashed line). For stronger diffusion, i.e.\ $E^{0.6}$, the primary spectra would have to be flatter than this. Such a behavior at Earth can either be reproduced by a broken power-law at the source, see e.g.\ \cite{biermann_prl2009,biermann_apj2010,biermann_apjl2010} or a broken power-law in the diffusion. 

    While the spectral behavior is debatable here, it becomes clear that with our numerical approach, we can reproduce the standard argument of SNRs being able to reproduce the cosmic ray energy budget. In the following, we will use realistic individual SNR spectra to test if those spectra that are observed today can represent the class of source that dominate the cosmic ray spectrum below the knee.

\subsection{Results simulating individual SNRs}
\begin{figure}
  \begin{center}
    \includegraphics[width=\textwidth]{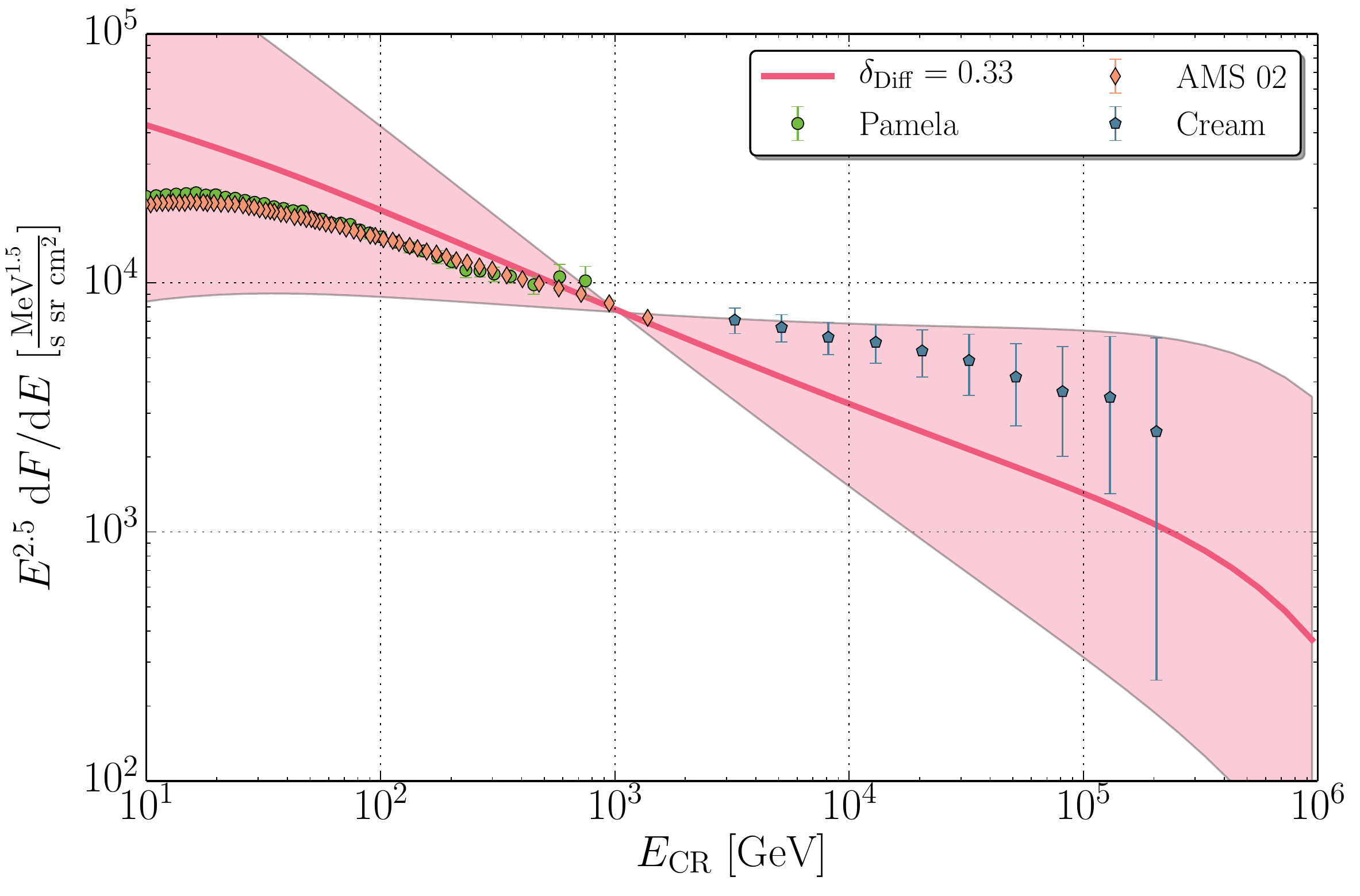}
  \end{center}
  \caption{\label{individual_1sigma_kolmogorov:fig}The CR flux in the large galaxy for simulated 20,000 SNRs, with the individual injection parameters taken from \cite{mandelartz2015}, considering a 1$\sigma$ error in the spectral index. The diffusion coefficient has been set to a Kolmogorov-type diffusion, i.e. $D\propto E^{0.33}$. Experimental data taken from CREAM \cite{2011ApJ...728..122Y}, PAMELA \cite{adriani2011} and AMS-01 \cite{2010arXiv1008.5051T}.}
    \end{figure}

\begin{figure}
  \begin{center}
    \includegraphics[width=\textwidth]{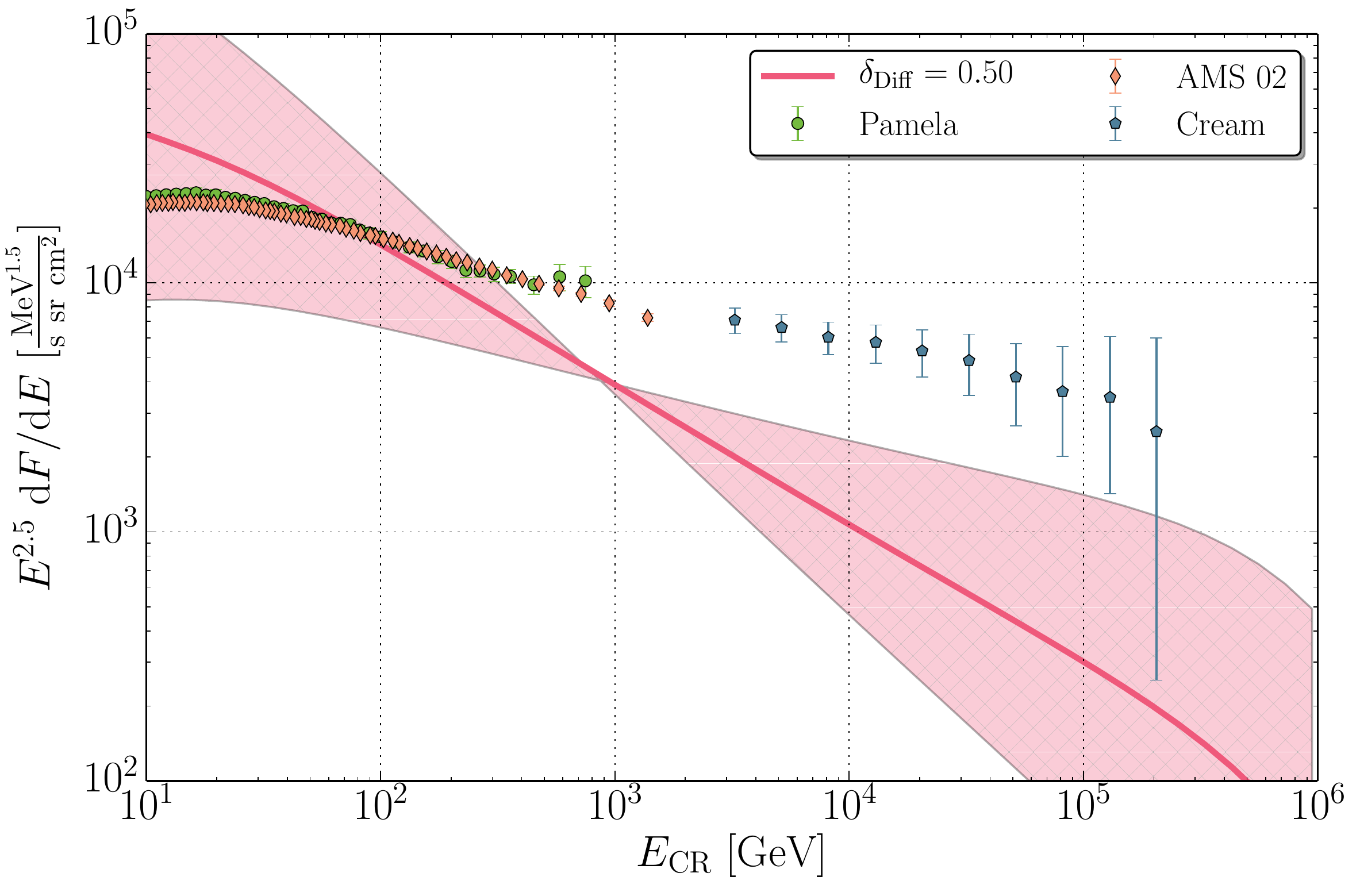}
  \end{center}
  \caption{\label{individual_1sigma:fig} The large galaxy with a 1$\sigma$ error in the individual spectral index. The diffusion coefficient has been chosen to be steeper than a Kolmogorov-type diffusion. Here, we use $D\propto E^{0.5}$. Experimental data taken from CREAM \cite{2011ApJ...728..122Y}, PAMELA \cite{2011Sci...332...69A} and AMS-01 \cite{2010arXiv1008.5051T}.
    }
  \end{figure}

\begin{figure}
  \begin{center}
    \includegraphics[width=\textwidth]{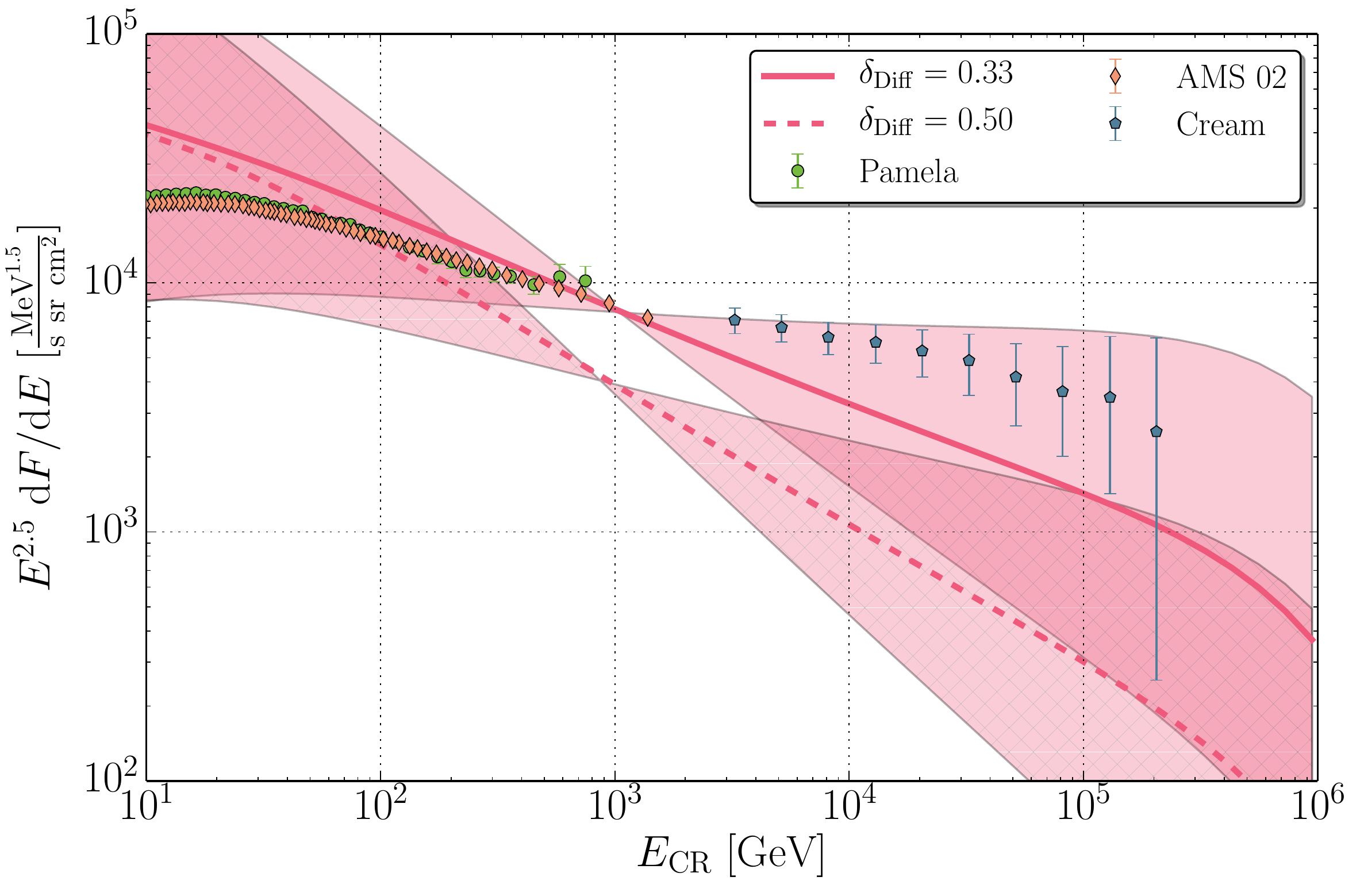}
  \end{center}
  \caption{\label{individual_diffusions:fig} The large galaxy examined with the Kolmogorov-type diffusion (solid, $\delta = 0.33$) and a steeper diffusion (dashed, $\delta=0.50$). Both simulations are shown along with the 1$\sigma$ error band of the spectral index. Experimental data taken from CREAM \cite{2011ApJ...728..122Y}, PAMELA \cite{2011Sci...332...69A} and AMS-01 \cite{2010arXiv1008.5051T}.}
  \end{figure}

In this section, we use those 21 SNRs with gamma-ray spectra that can be explained by hadronic interactions, as described in section \ref{method:sec}. Figure \ref{individual_1sigma_kolmogorov:fig} shows the results for a large galaxy with a Kolmogorov-type diffusion coefficient, i.e.\ $D\propto E^{0.33}$. Error bands show the 1$\sigma$ interval representing the uncertainty of the spectral index derived from the gamma-ray spectra. The cosmic ray data lie within these uncertainties and can be explained by these individual SNR spectra. Generally, the spectrum is somewhat steeper than what is observed on large scales. It is obvious, however, that uncertainties are still large and that future data with higher precision will have to confirm this result. Figure \ref{individual_1sigma:fig} shows the same simulation, but with a diffusion coefficient $D\propto E^{0.5}$. Here, low-energy data, i.e.\ below 10~TeV are well explained, but the high-energy tail cannot be reproduced. Figure \ref{individual_diffusions:fig} shows the result for the two different diffusion coefficients and the large galaxy setting to summarize the main results. These first results indicate that a Kolmogorov-like diffusion coefficient is well-suited to explain the cosmic ray flux, while a rather strong energy dependence of $E^{0.5}$ already leads to a spectrum that is too steep.


\subsection{Discussion of uncertainties}
The main uncertainties in this calculation are the following:
\begin{itemize}
\item We assume the number of sources actually contributing to the spectra at a given point of time to be $\sim 100$. This number could be larger, as close to $300$ radio emitting SNRs have been detected in the Milky Way so far. However, we use this low value as it is expected that only the brightest radio SNR are able to accelerate particles to extreme energies.
\item We use all gamma-ray detected SNRs that have the potential to be of hadronic origin. \cite{mandelartz2015} found that 21 out of 24 SNR gamma-ray spectra can be fitted hadronically, but it is not certain that all of these are dominated by $\pi^{0}$ decays.
\item The gamma-ray spectra are given between some GeV and $10$~TeV in energy, corresponding to cosmic ray energies of $10$~GeV to $100$~TeV. As the cosmic ray knee lies at about $1$~PeV, those spectra that do not show a cutoff up to $100$~TeV have to be extrapolated. The low-energy part, i.e.\ below $10$~GeV, cannot be described properly and is not subject of this investigation.
  \item The current simulation approach predicts the average CR observable, e.g.\ $<dF/dT>$, but not the proper corresponding variance, e.g.\ $Var(dF/dT)$. However, the variance or alternative statistical measures are needed to fully quantify whether the aforementioned discrepancy of $<dF/dT>$ is statistically significant.
    As more than those 21 known SNRs are expected to contribute to the CR flux the variance calculate here would presumably be an upper limit for the true variance.  In future investigations, we plan to remove this uncertainty by means of changing the method. This will be described in detail in Section \ref{discussion:sec}.

\end{itemize}
In the future, with HAWC and CTA data available, it will be possible to draw even stronger conclusions in particular what concerns the contribution up to the knee. At this point, our result rather has to be considered as an upper limit: all physics uncertainties have been considered in a way that the maximum possible result is received. In particular, all SNRs that can possibly be fit hadronically have been used and all spectra without a detected cutoff have been extrapolated with an assumed cutoff at the knee, i.e.\ at $1$~PeV. Thus, the conclusions of this paper are rather referring to an upper limit rather than a precise flux estimate.

\subsection{Conclusions}
We simulate the propagation of cosmic rays from individual SNRs in the Galaxy. We assume that all SNRs have spectra that are represented from a sample of 21 SNRs with measured gamma-ray spectra that can be fit hadronically. The aim of this paper is to start using gamma-ray data to investigate if SNRs can be the sources of cosmic rays below the knee. The uncertainties described above do not allow for a detailed comparison of the measured spectrum with simulations, but they give first evidence of the possible contribution by gamma-ray emitting SNRs.

The cosmic ray spectrum is within errors well-described by a Kolmogorov-type diffusion. However, if only a small fraction of the gamma-ray detections actually have a hadronic origin, it will be difficult to explain the entire spectrum by using SNRs. The same is true if several SNRs that do not show a cutoff in their spectrum yet do have an early cutoff. In that case, SNRs would fail to explain the high-energy part of the spectrum.

Stronger diffusion, i.e.\ $\delta = 0.5$ fails to explain the high-energy component of the spectrum as the total spectrum becomes too steep. So, even in the most optimistic scenario, it becomes difficult to explain the detected spectrum by SNRs.

The fact that only weak diffusion can describe the detected cosmic ray spectrum even in the optimistic scenario has further consequences. It implies that there is not much room for convection: in our simulation, we neglect any convective effects. Including a convective outflow in the calculations would remove parts of the energy budget coming from the sources, as some of the energy is carried out of the Galaxy. That means the simulated spectrum at Earth is expected to be even lower when convection effects are included. These results therefore indicate that convection cannot play a major role in the transport of cosmic rays. Further investigations are necessary to confirm these first results.




\section{Discussion and Outlook \label{discussion:sec}}

\subsection{Discussion}
We consider this paper as a first proof of concept that gamma-ray data can be used in the future to try to constrain the cosmic ray energy budget from supernova remnants. At this point, the data only allow for an upper limit calculation: the central conclusion is that the energy budget of the observed cosmic ray flux can be matched by the population of gamma-ray emitting SNRs. , they {\bf fail to explain the detected flux if...}
\begin{itemize}
\item ... several of those sources without gamma-ray detected cutoff actually cuts off earlier than in the knee region;
\item ... many of the gamma-ray signatures are lepton-dominated, i.e.\ produced by electrons rather than cosmic rays;
  \item ... diffusion is significantly stronger than the Kolmogorov-case ($\delta =0.33$).
  \end{itemize}
Future investigations will have to confirm these conclusions. In particular, data from CTA and HAWC will help to extend the measurements up to cosmic ray knee energies.

It is still interesting to see that, although gamma-ray measurements show very diverse spectra, the result matches the observed cosmic ray budget very well within the uncertainties.

One further conclusion is that that a cosmic ray wind is not needed to explain the data: without including convection, the energy budget can be well-reproduced. The introduction of convective effects would reduce the energy budget observed at Earth and thus possibly lead to the underestimation of the cosmic ray flux.

\subsection{Outlook}

One thing that we want to improve in the future is a better quantification of the variance of the method. To quantify the agreement between the predicted CR observables statistically, a measure of the variance is needed. The variance of the CR observables could be calculated and presented here along with the mean value. However, this variance has a limited statistical interpretation in the current simulation approach. It merely measures the spread of the simulated CR observable which is induced by randomly selecting positions for the 21 SNR in the Galaxy. In particular, this variance would decrease if the number of SNRs would be increased.  It is expected that the variance of those 21 known SNRs pose an upper limit of the true variance. It would be desirable, in particular as soon as gamma-ray data allow for a more precise calculation of the CR spectra at the source, to have precise knowledge of the variance.
The calculation of the true variance would require to include temporal aspects of the SNRs - such as their production rate and lifetime - into our simulation approach. This could be done as a generalization of our procedure to include spatial aspects via the selection random positions and mapping of SNRs to the closest grid point on the spatial simulation grid \cite{nils_ecrs2014}. As similar issues are already addressed in the GALPROP manual ~\cite{strong_galprop_2011,Galprop_Web_Standford} (see chapter 6), an implementation including the temporal aspects of SNRs should be feasible.

On longer terms we intend to perform Monte-Carlo (MC) simulations of the propagation of Galactic CRs. As a basis, we suggest to use the publicly available CRPropa MC-framework to study the propagation of UHECRs in extragalactic environments~\cite{2013APh....42...41K}. Especially the redesigned object orientated structure of the upcoming version 3.0 of CRPropa seems to allow for an easy extensions for the propagation of Galactic CRs ~\cite{2013arXiv1307.2643A}. With today's computer technologies, a Monte Carlo treatment of Galactic CRs down to T$\sim$10-100~TeV seems possible with reasonable run times. For lower energies it may be sufficient to switch to a diffusive approximation to avoid the time intensive numerical solution of the equation of motion in the Galactic magnetic field. This method has two central advantages with respect to the approach of solving the transport equation:
\begin{itemize}
\item As single particles are propagated, it is possible to follow each individual trajectory.
  \item The method allows for the implementation of arbitrary Galactic field models. This means that it is not necessary to assume a diffusion scalar, but that the particles can travel through a realistic magnetic field. By comparing different models, the effect of the magnetic field on the propagation and detection at Earth is possible. It can even be possible to derive the diffusion tensor using this method.
\end{itemize}



\section*{Acknowledgments}
 We warmly thank Peter Biermann, John Black, Ralf-J\"urgen Dettmar,  Carmelo Evoli, Horst Fichtner, Philipp Graeser, Francis Halzen, Matthias Mandelartz, Florian Schuppan and Andy Strong for intensive and fruitful discussions.
We
acknowledge the support from the DFG for the project B4 {\it Transport of cosmic rays from supernova remnants through the Galactic Magnetic Field} within the research unit ``Instabilities, turbulence and transport in
cosmic magnetic fields'' (FOR1048), from the Research Department of Plasmas with Complex Interactions (Bochum) and from the MERCUR-funded Ruhr Astroparticle-Plasma Physics centrum, RAPP centrum (St-2014-0040). 

\bibliographystyle{elsarticle-num}
\bibliography{lib_transport}

\begin{thebibliography}{10}
\expandafter\ifx\csname url\endcsname\relax
  \def\url#1{\texttt{#1}}\fi
\expandafter\ifx\csname urlprefix\endcsname\relax\def\urlprefix{URL }\fi
\expandafter\ifx\csname href\endcsname\relax
  \def\href#1#2{#2} \def\path#1{#1}\fi

\bibitem{hess1912}
V.~F. {Hess}, Phys.~Z. 13 (1912) 1084.

\bibitem{gaisser1991}
T.~{Gaisser}, {Cosmic Rays and Particle Physics}, Cambridge University Press,
  1991.

\bibitem{stanev2003}
T.~Stanev, High-energy cosmic rays, 2nd Edition, Springer Praxis
  Books/Astronomy and Planetary Sciences, 2010.

\bibitem{blumenthal_gould1970}
G.~R. {Blumenthal}, R.~J. {Gould}, Reviews of Modern Physics 42 (1970) 237.

\bibitem{schlickeiser2002}
R.~{Schlickeiser}, {Cosmic Ray Astrophysics}, Springer, Berlin, 2002.

\bibitem{abdo_w44}
A.~A. {Abdo}, et~al., Science 327 (2010) 1103.

\bibitem{adbo_ic443}
A.~A. {Abdo}, et~al., Astroph.~J. 712 (2010) 459.

\bibitem{tevcat}
{TeV Catalog}, \url{http://tevcat.uchicago.edu/} (Jan 2015).

\bibitem{halzen_klein2010}
F.~{Halzen}, S.~R. {Klein}, Review of Scientific Instruments 81~(8) (2010)
  081101.

\bibitem{icecube2013}
M.~{Aartsen}, et~al., Science 342.

\bibitem{icecube2014}
M.~G. {Aartsen}, et~al., Phys.~Rev.~Lett. 113~(10) (2014) 101101.

\bibitem{kistler_beacom2006}
M.~D. {Kistler}, J.~F. {Beacom}, Phys.~Rev.~D 74~(6) (2006) 063007.

\bibitem{ahlers_murase2014}
M.~{Ahlers}, K.~{Murase}, Phys.~Rev.~D 90~(2) (2014) 023010.

\bibitem{neronov2014}
A.~{Neronov}, D.~{Semikoz}, C.~{Tchernin}, Phys.~Rev.~D 89~(10) (2014) 103002.

\bibitem{winter_galactic2014}
J.~C. {Joshi}, W.~{Winter}, N.~{Gupta}, Mon.~Not.~Roy.~Astron.~Soc. 439 (2014)
  3414.

\bibitem{kachelriess2014}
M.~{Kachelrie{\ss}}, S.~{Ostapchenko}, Phys.~Rev.~D 90~(8) (2014) 083002.
\newblock \href {http://arxiv.org/abs/1405.3797} {\path{arXiv:1405.3797}}.

\bibitem{mandelartz2015}
M.~{Mandelartz}, J.~{Becker Tjus}, Astropart.~Phys. 65 (2015) 80.

\bibitem{icecubegen2_whitepaper}
M.~{Aartsen}, et~al., {IceCube-Gen2: A Vision for the Future of Neutrino
  Astronomy}, arXiv:1412.5106.

\bibitem{haungs2011}
A.~{Haungs}, Astrophysics and Space Sciences Transactions 7 (2011) 295.

\bibitem{blasi2013}
P.~{Blasi}, Astron. \& Astroph. Rev. 21 (2013) 70.
\newblock \href {http://arxiv.org/abs/1311.7346} {\path{arXiv:1311.7346}}.

\bibitem{wolfendale2014}
A.~{Wolfendale}, A.~{Erlykin}, Astropart.~Phys. 53 (2014) 115.
\newblock \href {http://dx.doi.org/10.1016/j.astropartphys.2013.01.009}
  {\path{doi:10.1016/j.astropartphys.2013.01.009}}.

\bibitem{ahn2010}
H.~S. {Ahn}, et~al., Astroph.~J.~Lett. 714 (2010) L89.

\bibitem{biermann_apj2010}
P.~L. {Biermann}, et~al., Astroph.~J. 725 (2010) 184.

\bibitem{ams02_protons_icrc2013}
S.~{Haino}, et~al.,
  \href{http://143.107.180.38/indico/contributionDisplay.py?contribId=1265&sessionId=3&confId=0}{{Precision
  measurement of the proton flux with AMS}}, International Cosmic Ray
  Conference, 2013.
\newline\urlprefix\url{http://143.107.180.38/indico/contributionDisplay.py?contribId=1265&sessionId=3&confId=0}

\bibitem{ams02_helium_icrc2013}
V.~{Choutko}, et~al.,
  \href{http://143.107.180.38/indico/contributionDisplay.py?contribId=1262&sessionId=3&confId=0}{{Precision
  measurement of the helium flux with AMS}}, International Cosmic Ray
  Conference, 2013.
\newline\urlprefix\url{http://143.107.180.38/indico/contributionDisplay.py?contribId=1262&sessionId=3&confId=0}

\bibitem{kascade_grande2013}
W.~D. {Apel}, et~al., Astropart.~Phys. 47 (2013) 54.

\bibitem{kascade_grande2014}
W.~D. {Apel}, et~al., Advances in Space Research 53 (2014) 1456.

\bibitem{aartsen_icetop2013}
M.~G. {Aartsen}, et~al., Phys.~Rev.~D 88~(4) (2013) 042004.

\bibitem{auger_composition_icrc2013}
D.~{Garc{\'i}a-G{\'a}mez}, et~al., {Observations of the longitudinal
  development of extensive air showers with the surface detectors of the Pierre
  Auger Observatory}, International Cosmic Ray Conference, 2013,
  arXiv:1307.5059.

\bibitem{strong_propagation_1998}
A.~W. Strong, I.~V. Moskalenko,
  \href{http://iopscience.iop.org/0004-637X/509/1/212}{Propagation of
  cosmic-ray nucleons in the galaxy} 509~(1)  212.
\newline\urlprefix\url{http://iopscience.iop.org/0004-637X/509/1/212}

\bibitem{dragon}
C.~{Evoli}, et~al., {Cosmic ray nuclei, antiprotons and gamma rays in the
  galaxy: a new diffusion model}, {JCAP} 10 (2008) 18.
\newblock \href {http://arxiv.org/abs/0807.4730} {\path{arXiv:0807.4730}},
  \href {http://dx.doi.org/10.1088/1475-7516/2008/10/018}
  {\path{doi:10.1088/1475-7516/2008/10/018}}.

\bibitem{kissmann2014}
R.~{Kissmann}, Astropart.~Phys. 55 (2014) 37.
\newblock \href {http://arxiv.org/abs/1401.4035} {\path{arXiv:1401.4035}},
  \href {http://dx.doi.org/10.1016/j.astropartphys.2014.02.002}
  {\path{doi:10.1016/j.astropartphys.2014.02.002}}.

\bibitem{strong_new_1998}
A.~W. Strong, I.~V. Moskalenko,
  \href{http://arxiv.org/abs/astro-ph/9807289}{New constraints on galactic
  cosmic-ray propagation}.
\newline\urlprefix\url{http://arxiv.org/abs/astro-ph/9807289}

\bibitem{abdo_milagro2007}
A.~A. {Abdo}, {(MILAGRO Coll.)}, et~al., Astroph.~J.~Lett. 658 (2007) L33.
\newblock \href {http://arxiv.org/abs/arXiv:astro-ph/0611691}
  {\path{arXiv:arXiv:astro-ph/0611691}}.

\bibitem{abdo_milagro2008}
A.~A. {Abdo}, {(MILAGRO Coll.)}, et~al., Astroph.~J. 688 (2008) 1078.
\newblock \href {http://arxiv.org/abs/0805.0417} {\path{arXiv:0805.0417}}.

\bibitem{abdo_extragalactic2010}
A.~A. {Abdo}, {(Fermi Coll.)}, et~al., Phys.~Rev.~Lett. 104~(10) (2010) 101101.
\newblock \href {http://arxiv.org/abs/1002.3603} {\path{arXiv:1002.3603}}.

\bibitem{strong2010}
A.~W. {Strong}, et~al., Astroph.~J.~Lett. 722 (2010) L58.
\newblock \href {http://arxiv.org/abs/1008.4330} {\path{arXiv:1008.4330}}.

\bibitem{buesching2008}
I.~{B{\"u}sching}, C.~{Venter}, O.~C. {de Jager}, Advances in Space Research 42
  (2008) 497.

\bibitem{ahlers2009}
M.~{Ahlers}, P.~{Mertsch}, S.~{Sarkar}, Phys.~Rev.~D 80~(12) (2009) 123017.
\newblock \href {http://arxiv.org/abs/0909.4060} {\path{arXiv:0909.4060}}.

\bibitem{erlykin2013}
A.~D. {Erlykin}, A.~W. {Wolfendale}, {Cosmic ray positrons from a local,
  middle-aged supernova remnant}, Astroparticle Physics 49 (2013) 23--27.
\newblock \href {http://arxiv.org/abs/1308.4878} {\path{arXiv:1308.4878}},
  \href {http://dx.doi.org/10.1016/j.astropartphys.2013.08.001}
  {\path{doi:10.1016/j.astropartphys.2013.08.001}}.

\bibitem{biermann_prl2009}
P.~L. {Biermann}, et~al., Phys.~Rev.~Lett. 103~(6) (2009) 061101.

\bibitem{pamela2009}
O.~{Adriani}, {(PAMELA Coll.)}, et~al., Nature 458 (2009) 607.

\bibitem{ams_positrons2013}
M.~{Aguilar}, et~al., Phys.~Rev.~Lett. 110~(14) (2013) 141102.

\bibitem{biermann_apjl2010}
P.~L. {Biermann}, et~al., Astroph.~J.~Lett. 710 (2010) L53.

\bibitem{finkbeiner_galactic_wind2010}
M.~{Su}, T.~R. {Slatyer}, D.~P. {Finkbeiner}, Astroph.~J. 724 (2010) 1044.
\newblock \href {http://arxiv.org/abs/1005.5480} {\path{arXiv:1005.5480}}.

\bibitem{amenomori_ta_anisotropy2006}
M.~{Amenomori}, et~al., Science 314 (2006) 439.

\bibitem{guillian_superk_anisotropy2007}
G.~{Guillian}, {SuperK.\ Coll.}, et~al., Phys.~Rev.~D 75~(6) (2007) 062003.

\bibitem{abdo_milagro_anisotropy2009}
A.~A. {Abdo}, et~al., Astroph.~J. 698 (2009) 2121.

\bibitem{abbasi2010}
R.~{Abbasi}, et~al., Astroph.~J.~Lett. 718 (2010) L194.

\bibitem{giacinti_sigl2012}
G.~{Giacinti}, G.~{Sigl}, Phys.~Rev.~Lett. 109~(7) (2012) 071101.

\bibitem{sveshnikova2013}
L.~G. {Sveshnikova}, O.~N. {Strelnikova}, V.~S. {Ptuskin}, Astropart.~Phys. 50
  (2013) 33.

\bibitem{pohl2013}
M.~{Pohl}, D.~{Eichler}, Astroph.~J. 766 (2013) 4.

\bibitem{biermann_anisotropie}
P.~L. {Biermann}, et~al., Astroph.~J. 768 (2013) 124.

\bibitem{gaggero2013}
D.~{Gaggero}, et~al., Phys.~Rev.~Lett. 111~(2) (2013) 021102.
\newblock \href {http://arxiv.org/abs/1304.6718} {\path{arXiv:1304.6718}}.

\bibitem{gaggero2014}
D.~{Gaggero}, et~al., Phys.~Rev.~D 89~(8) (2014) 083007.
\newblock \href {http://arxiv.org/abs/1311.5575} {\path{arXiv:1311.5575}}.

\bibitem{werner2015}
M.~{Werner}, et~al., Astropart.~Phys. 64 (2015) 18.
\newblock \href {http://arxiv.org/abs/1410.5266} {\path{arXiv:1410.5266}}.

\bibitem{drury2014}
L.~{O'C.~Drury}, Invited talk at CRISM2014\href
  {http://arxiv.org/abs/arXiv:1412.1376} {\path{arXiv:arXiv:1412.1376}}.

\bibitem{diehl2006}
R.~{Diehl}, et~al., nat 439 (2006) 45.
\newblock \href {http://arxiv.org/abs/astro-ph/0601015}
  {\path{arXiv:astro-ph/0601015}}.

\bibitem{yanasak2001}
N.~E. {Yanasak}, et~al., Astroph.~J. 563 (2001) 768.

\bibitem{Galprop_Web_Standford}
\href{http://galprop.stanford.edu/}{Galprop v54} (2014).
\newline\urlprefix\url{http://galprop.stanford.edu/}

\bibitem{schlickeiser_jenko2010}
R.~{Schlickeiser}, F.~{Jenko}, Journal of Plasma Physics 76 (2010) 317.

\bibitem{green2014}
D.~A. {Green}, Bulletin of the Astronomical Society of India 42 (2014) 47.
\newblock \href {http://arxiv.org/abs/1409.0637} {\path{arXiv:1409.0637}}.

\bibitem{1998ApJ...504..761C}
G.~L. {Case}, D.~{Bhattacharya}, {A New {$\Sigma$}-D Relation and Its
  Application to the Galactic Supernova Remnant Distribution}, Astroph.~J. 504
  (1998) 761--772.
\newblock \href {http://arxiv.org/abs/astro-ph/9807162}
  {\path{arXiv:astro-ph/9807162}}, \href {http://dx.doi.org/10.1086/306089}
  {\path{doi:10.1086/306089}}.

\bibitem{2009BASI...37...45G}
D.~A. {Green}, {A revised Galactic supernova remnant catalogue}, Bulletin of
  the Astronomical Society of India 37 (2009) 45--61.
\newblock \href {http://arxiv.org/abs/0905.3699} {\path{arXiv:0905.3699}}.

\bibitem{hawc}
A.~U. {Abeysekara}, et~al., Astropart.~Phys. 50 (2013) 26.

\bibitem{cta}
B.~S. {Acharya}, et~al., Astropart.~Phys. 43 (2013) 3.

\bibitem{strong_galprop_2011}
A.~W. Strong, et~al.,
  \href{http://130.79.128.5/ftp/0/aa/534/A54/galprop\_v54.pdf}{{GALPROP}
  version 54: Explanatory supplement} (2011).
\newline\urlprefix\url{http://130.79.128.5/ftp/0/aa/534/A54/galprop\_v54.pdf}

\bibitem{cox1972}
D.~P. {Cox}, Astroph.~J. 178 (1972) 159.

\bibitem{2011ApJ...728..122Y}
Y.~S. {Yoon}, et~al., {Cosmic-ray Proton and Helium Spectra from the First
  CREAM Flight}, APJ 728 (2011) 122.
\newblock \href {http://arxiv.org/abs/1102.2575} {\path{arXiv:1102.2575}},
  \href {http://dx.doi.org/10.1088/0004-637X/728/2/122}
  {\path{doi:10.1088/0004-637X/728/2/122}}.

\bibitem{adriani2011}
O.~{Adriani}, et~al., Science 332 (2011) 69.

\bibitem{2010arXiv1008.5051T}
{The AMS-01 Collaboration}, {Relative Composition and Energy Spectra of Light
  Nuclei in Cosmic Rays. Results from AMS-01}, ArXiv e-prints\href
  {http://arxiv.org/abs/1008.5051} {\path{arXiv:1008.5051}}.

\bibitem{2011Sci...332...69A}
O.~{Adriani}, et~al., {PAMELA Measurements of Cosmic-Ray Proton and Helium
  Spectra}, Science 332 (2011) 69--.
\newblock \href {http://arxiv.org/abs/1103.4055} {\path{arXiv:1103.4055}},
  \href {http://dx.doi.org/10.1126/science.1199172}
  {\path{doi:10.1126/science.1199172}}.

\bibitem{nils_ecrs2014}
N.~{Nierstenh\"ofer}, et~al., {Galactic Propagation of Cosmic Rays from
  Individual Supernova Remnants}, Proceedings to be published IOP conference
  series (September 2014).

\bibitem{2013APh....42...41K}
K.-H. {Kampert}, et~al., {CRPropa 2.0 - A public framework for propagating high
  energy nuclei, secondary gamma rays and neutrinos}, {Astrop.\ Phys.} 42
  (2013) 41.
\newblock \href {http://arxiv.org/abs/1206.3132} {\path{arXiv:1206.3132}},
  \href {http://dx.doi.org/10.1016/j.astropartphys.2012.12.001}
  {\path{doi:10.1016/j.astropartphys.2012.12.001}}.

\bibitem{2013arXiv1307.2643A}
R.~{Alves Batista}, et~al., {CRPropa 3.0 - a Public Framework for Propagating
  UHE Cosmic Rays through Galactic and Extragalactic Space}, ArXiv
  e-prints\href {http://arxiv.org/abs/1307.2643} {\path{arXiv:1307.2643}}.

\end{thebibliography}

\end{document}